# The magnetic moment enigma in Fe-based high temperature superconductors

Norman Mannella
Department of Physics and Astronomy, University of Tennessee – Knoxville

**ABSTRACT**

The determination of the most appropriate starting point for the theoretical description of Fe-based materials hosting high temperature superconductivity remains among the most important unsolved problem in this relatively new field. Most of the work to date has focused on the pnictides, with LaFeAsO, $BaFe_2As_2$ and LiFeAs being representative parent compounds of three families known as 1111, 122 and 111, respectively. This Topic Review examines recent progress in this area, with particular emphasis on the implication of experimental data which have provided evidence for the presence of electron itinerancy and the detection of local spin moments. In light of the results presented, the necessity of a theoretical framework contemplating the presence and the interplay between itinerant electrons and large spin moments is discussed. It is argued that the physics at the heart of the macroscopic properties of pnictides Fe-based high temperature superconductors appears to be far more complex and interesting than initially predicted.

## 1. INTRODUCTION

The study of iron (Fe)-based materials hosting high temperature superconductivity (HTSC) continues to receive a great deal of interest in the community. The initial phases of the discovery of HTSC in Fe-based high temperature superconductors (Fe-HTSC) bear some similarities with the discovery of HTSC in Cu oxide materials (cuprates) by Bednorz and Muller in 1986 [1]. The fact that HTSC could be hosted in materials whose parent compounds are oxide insulators certainly came as a surprise. Similarly, the discovery of HTSC in Fe-HTSC was surprising in light of the fact that common wisdom suggested that the strong local moment carried by the Fe atom was detrimental for superconductivity. In fact, instances of superconductivity with low (<10 K) superconducting critical temperatures ($T_C$) in compounds with non-magnetic Fe have been known for quite some time [2,3,4,5]. Metallic Fe itself, which is ~~even~~ magnetic, is a superconductor with $T_C \approx 1.8$ K under pressure of 20 GPa [6].

The first class of Fe-based compounds found to host HTSC is that of Pnictide oxides, i.e. quaternary rare earth transition metals containing pnictogens, the chemical elements found in group 15 of the Periodic Table such as Phosphorous (P), Arsenic (As) and Antimony (Sb). Referred to as the "1111" compounds (cf. Fig. 1), pnictide oxides have formula unit RETPnO (RE=Rare Earth, T=Transition metal, Pn=Pnictogen). The first report of RETPnO compounds was provided by the Jeitschko group back in 1996 [7]. The same group reported eighteen quaternary arsenides such as LaFeAsO a few years later [8]. Superconductivity with $T_C = 4$ K was first reported in LaFePO phosphide oxides and pushed to $T_C = 7$ K upon F doping [9].

The attention of the community towards Fe-based superconductors was immediately polarized after the discovery of superconductivity with $T_C = 26$ K reported in F-doped LaFeAsO by Kamihara in Feb. 2008 [10], which signals the start of the field of HTSC in Fe-based materials. Since then, the field progressed at an incredible fast pace, due to the experience that the community had gained by studying cuprates. By April 2008, $T_C$ was increased up to ≈ 50 K by replacing La with other rare earth elements such as Ce, Pr, Nd, Sm and Gd [11]. Superconductivity with $T_C = 38$ K was reported in $Ba_{1-x}K_xFe_2As_2$ compounds with the $ThCr_2Si_2$ structure in May 2008 [12]. The discovery of HTSC in compounds with the $ThCr_2Si_2$ structure, now referred to as the "122" compounds, was particularly significant since it indicated that HTSC was not hosted exclusively in oxide materials. By July of the same year, superconductivity with $T_C = 18$ K was found in $Li_xFeAs$ [13], while $Fe_{1+x}Se$ exhibited superconductivity with $T_C = 8$ K [14]. The critical temperature can be increased dramatically by Te substitution [15], or even more by pressure up to 37 K [16,17], providing two new structure types, the "111" and the chalcogenides FeCh (Ch = S, Se, Te) "11" families. It is important to remark that compounds of the 11 family have a simpler structure than the pnictides since there are no atoms in between the FeSe layers (cf. Fig. 1). Together with these four structures, one has to consider more recent developments consisting of the discovery of HTSC in other systems. These include the "21311" compounds $Sr_2MO_3FePn$ (M = Sc, V and Cr,) with $T_C = 17$ K (with M = Sc and Pn = P) [18], and $T_C = 37$ K (with M = V and Pn) [19], a variant of the 21311 structure as found in $Ca_2(Mg_{0.25}Ti_{0.75})_{1.5}O_4FeAs$ with $T_C = 47$ K [20], the alkali metal iron selenide $A_xFe_{2-y}Se_2$ (A=alkali metal) [21], with typical alkali metal elements intercalated in between the FeSe layers, also referred to as 122 with defect structure and labeled 122*, and the most recent discovery of HTSC ($T_C = 65$ K) on a single sheet of FeSe grown on $SrTiO_3$ [22,23,24,25,26].

A detailed survey of the properties of these different families of Fe-HTSC can be found in many reviews of this new field. Some offer a comprehensive overview, including the beginning [27,28,29,30] and more recent work and perspectives [31,32,33,34], while others emphasize more specific aspects of the field such as magnetic properties [35,36], theoretical approaches [37], 122 materials [38], iron chalcogenides [39] and alkali metal iron selenides [40].

As emphasized by Stewart [32], despite differences in their properties, all of the families of Fe-HTSC compounds share many commonalities. No matter what the structure is (i) the basic structural building blocks are square planar nets of Fe atoms arranged in a tetrahedral environment, with bond angles and positions of the Pn/Ch atom above Fe in the tetrahedra displaying a correlation with $T_C$; (ii) the Fe-HTSC are multi-orbital systems with a complex Fermi surface consisting of different bands originating from the hybridization of the Fe *d* orbitals; (iii) the Fe-HTSC are unconventional superconductors in light of the elimination of conventional BCS-like pairing mechanisms [41,42,43]; (iv) with a few exceptions (such as LiFeAs and FeSe exhibiting no magnetism), the Fe atom is found to be magnetic in many parts of the phase diagrams. The parent compounds exhibit a long range antiferromagnetic order which is suppressed with concomitant emergence of superconductivity as additional carriers are introduced into, or pressure is applied to the system [31].

The determination of the most appropriate starting point for a theoretical description of the Fe-HTSC remains among the most important unresolved problems in this field. In the context of a general view of HTSC in Fe-based materials, the previous statement is arguably true even more so today as compared to the years immediately following the start of the field [34]. The two fundamental topics of discussion are the role of electron correlations and the microscopic origin of magnetism.

Up to 2010, most of the work concentrated on the 1111, 122 and 111 pnictide families. Compounds belonging to these families are uniformly metallic throughout their doping/pressure phase diagrams. Many experiments indicated that the Fe-HTSC are capable of hosting HTSC without the signatures of strong local Mott-Hubbard type correlations that characterize cuprate HTSC, and confirmed many predictions of electronic band structure calculations [44,45,46,47,48,49], such as an itinerant Fe d-band character, a high density of states at the Fermi level, and the presence of Fermi surfaces composed of nearly cylindrical hole and electron pockets at the zone center and zone corners, respectively. The topology of the Fermi surface (FS) was believed to play a crucial role in the physics of the Fe-HTSC because the nesting of the electron and hole pockets leads to an enhancement of the particle-hole susceptibility, with the likely possibility of inducing spin-density-wave (SDW) order at the in-plane antiferromagnetic (AFM) wave vector $Q_{AF} = (\pi,0)$ (in the 2-Fe unit cell) with a collinear spin structure [50], much like the FS-nesting-induced SDW in pure Chromium [51]. The predicted AFM spin structure has been confirmed in the 1111, 122 and 111 families by neutron scattering experiments in LaFeAsO [52], $BaFe_2As_2$ [53] and NaFeAs [54] compounds, including the spin resonance, (a peak in the imaginary part of the dynamic susceptibility).

Discussions about the degree of Hubbard-type electron correlation in the Fe-HTSC took place since the start of the field. Different theoretical studies have reached opposite conclusions regarding the magnitude of the on-site Coulomb repulsion, denoted by the Hubbard parameter *U*, thus identifying the Fe-HTSC as weakly, moderately, or even strongly correlated systems [55,56,57,58,59,60,61]. It has been proposed that electron correlations could be sufficiently strong to produce "incipient" Mott physics [62,63], where localized and itinerant electrons may be equally important for a correct description of the Fe-HTSC [64,65]. It has been discussed how the $s_{\pm}$ pairing symmetry can be derived in a *t-J* model-framework [66,67] and, more recently, how the AFM state evolves smoothly from weak to strong coupling, suggesting that the physics of the pnictides could also be described by concepts markedly different from frameworks encompassing weak-coupling nesting [34,68].

Comparisons to the strongly correlated cuprates were initially widespread. The fact that pnictide materials are uniformly metallic throughout their doping/pressure phase diagrams, and that conventional density functional methods, which typically fail for correlated systems, were shown to capture many of the electronic properties, shaped the belief that AFM in the Fe-HTSC materials originated from FS nesting of itinerant electrons, and that Coulomb correlations did not play a significant role in the magnetism and HTSC.

As pointed out in a couple of recent reviews [34,40], this perspective may change in light of the provision of new experimental and theoretical results following in particular the study of the more recently discovered chalcogenides FeCh (Ch = S, Se, Te) "11" and alkali metal iron selenide $A_xFe_{2-y}Se_2$ (A=alkali metal) 122* families. The study of these new families, both of which have been recently reviewed [39,40], has revamped the importance of electron correlations for the Fe-HTSC.

Typical examples of the "11" compounds are FeSe and Te-doped FeSe with formula $FeTe_xSe_{1-x}$. The magnetism in $FeTe_{0.35}Se_{0.65}$ appears to be well described by a local picture, with large local spin moments on the Fe site being detected with inelastic neutron scattering experiments [69], and with angle resolved photoemission measurements suggesting that the normal state of $FeTe_xSe_{1-x}$ is electronically more correlated than that of the pnictides [70]. In addition to the fact that by now their critical temperatures are comparable to those of the pnictides, the interest in $A_xFe_{2-y}Se_2$ alkaline iron selenides is motivated by the difference of certain properties with respect to the pnictides. These include the occurrence of insulating states in different compositions, the absence of hole pockets in the FS, the presence of large local spin moments, and the presence of phase separation [40]. The $A_xFe_{2-y}Se_2$ selenides cannot be considered weakly correlated materials, in light of the intermediate values of the Hubbard repulsion *U* which have been found necessary for describing some of these properties such as the large local moments [40]. The absence of hole pockets dismisses mechanisms related to FS nesting between electron and hole pockets as necessary for HTSC in Fe-HTSC. Moreover, in the alkali metal iron selenides the symmetry of the superconducting order parameter appears constrained to options that are different from the $s_{\pm}$ proposed for the pnictides [40].

These results may cause a shift in the emerging paradigms for the whole field of Fe-HTSC, including the possible ways of viewing a comparison between Fe-HTSC and the cuprates. In both classes of materials, HTSC emerges in close proximity to a long-range-ordered AFM ground state, suggesting that magnetic fluctuations and other unconventional pairing mechanisms that do not rely on phonons are responsible for HTSC (cf. Fig. 2). Nonetheless, it is not clear at the moment what is the correct picture embracing the whole field of Fe-HTSC. Although they have the same magnetic groundstate, pnictides and alkali metal iron selenides could be considered different classes of materials, with different pairing mechanisms. Alternatively, one may think that a unifying principle could be operative in both families of compounds, despite differences in the strength of the Hubbard *U*, the nature of magnetism (itinerant vs. local), and the nature of the parent compounds (metallic vs. insulating) [34,71].

The microscopic origin of magnetism has been another point of debate. The opportunity of studying HTSC and its relation to magnetism in a wide range of magnetic element-based materials is certainly one of the benefits provided by the discovery of Fe-HTSC materials. The microscopic origin of magnetism in the Fe-HTSC and its role related to HTSC continues to be a subject of great interest in the community. One obvious reason is that the proximity of the superconducting state to magnetically ordered states suggests that the superconducting pairing mechanism may be related to the coexistent magnetism in the phase diagram. An equally compelling reason is that the microscopic origin of magnetism in Fe-HTSC is intimately tied to the electronic structure and the degree of electron correlations existing in these materials. - Both need to be understood in order to provide a correct theoretical description of the mechanisms underpinning the macroscopic properties, including HTSC.

While intermediate values of the Hubbard *U* provide a rationale for the occurrence of local moments found in chalcogenides and alkali metal iron selenides, in the pnictides the microscopic origin of magnetism appears to be more enigmatic. Although itinerant SDW and FS nesting have been usually taken as valid mechanisms for describing the magnetic states, recent spectroscopic results discussed in the next sections have revealed the presence of local spin moments in the paramagnetic phase with magnitude comparable to, or larger than, the values reported by neutron scattering in the low temperature ordered phases. The presence of local moments in the paramagnetic phase is incompatible with theories relying exclusively on FS instabilities, e.g. SDW, since in these frameworks the processes of moment formation and long range ordering formation should occur concomitantly at the antiferromagnetic ordering temperature or Neel temperature $T_N$. At the same time, the pnictides exhibit an itinerant electron character which is incompatible with the occurrence of strong local Mott-Hubbard type correlations that characterize for example the cuprates. Several theoretical works have indicated that perspectives viewing the electrons as completely localized or completely itinerant are inadequate for a correct description of the pnictides. In particular, works based on dynamical mean field theory have indicated the presence of strong on-site exchange interactions (Hund's coupling) which, via spin-fluctuations that couple to the *d*-electrons, can provide a rationale for the occurrence of large local moments and renormalizations and shifts in spectral weight within the Fe *d*-bands while maintaining an itinerant electron character.

This Topic Review discusses specifically the problem of the coexistence of itinerant electrons and local moments in the pnictides. The emphasis is on experimental data which have provided evidence for the presence of electron itinerancy and the detection of local moments. In light of the results presented, the necessity of a theoretical description capable of including these seemingly contrasting characteristics is discussed. Finally, a perspective on the importance of magnetic correlations in these compounds and the relevance to this problem of the short time scale typical of electron dynamics is discussed.

The organization of the paper is as follows: the degree of Coulomb-type electron correlations in pnictides will be discussed first, followed by a review of the results supporting the existence of local moments. In order to elucidate the importance of electron dynamics for understanding the formation of local moments in itinerant systems, the case of ferromagnetism of metallic Fe is discussed next. Lastly, considerations following the discussion of magnetism in metallic Fe will be shown to be relevant in order to provide a rationale for the coexistence of itinerant electrons and local moments in the pnictides.

## 2. COULOMB-TYPE CORRELATIONS IN PNICTIDES

The emergence of superconductivity in close proximity to a long-range-ordered AFM ground state and the similarity of the phase diagrams, shown in Fig. 2, initially suggested a close resemblance between the Fe-pnictides and the cuprates [72,73,74,75]. The degree of correlation in the Fe-HTSC and the closely intertwined microscopic nature of the magnetism has been widely discussed and debated since their discovery. The reason for this debate can be tracked back to the provision of different experimental results, which will be discussed below, providing seemingly contrasting information about the degree of electron itinerancy or localization. This had profound repercussions on the proposed mechanisms responsible for the microscopic origin of magnetism, and hence the enigma referred to in the title. According to mainstream viewpoints, the itinerant electron character suggests that the magnetic ordering is driven by FS nesting between the hole and electron pockets. On the contrary, the presence of local moments suggests that the magnetic ordering is caused by the interactions among the moments themselves.

To establish the degree of Coulomb-type electron correlations one often consider the magnitude of the Hubbard *U* relative to the electron bandwidth *W*, i.e. the ratio $U/W$. The *U* is the parameter that controls the on-site Coulomb interaction between localized electrons in the Hubbard model, a version of which is imported into density functional theory as "LDA+U", or in dynamical mean field theory calculations [76]. It is defined as the energy involved in the excitation of a *d*-electron from a metal ion of configuration $d^n$ onto another distant metal ion with the same configuration, that is, the energy involved in the charge fluctuation $d^n + d^n + U = d^{n+1} + d^{n-1}$. The value $U/W \simeq 1$ usually marks the separation between weakly and strongly correlated systems. Values of *U* have been found to range considerably, from $U \leq 2$ [55], to $2.2 \leq U \leq 3.3$ [77,78,79,80 ], to $U = 4$ eV [60,81], thus identifying the Fe-HTSC as being weakly, moderately, or even strongly correlated systems, in light of typical bandwidth values $W \simeq 4$ - 6 eV.

Different values of $U$ were proposed on the basis of comparisons or fits to experimental data available in the literature. The review by Johnston examines in great detail the early experiments and their interpretation [31]. It needs to be stressed how some of the first experimental results were not trustworthy because high quality samples were not yet available. For example, Fe-HTSC have been characterized as “bad metals” on the basis of the low electrical conductivity at room temperature as compared to that of simple metals like Copper. Here “bad metal” indicates that the calculated mean free path $l$ for conduction electrons is comparable to or less than the inter-atomic distance or, equivalently, $k_F l \leq 1$, where $k_F$ denotes the Fermi wavevector. In this case, the wavevector is no longer a good quantum number, and the electron excitations are incoherent, with a small value of the quasiparticle weight at the Fermi level [31]. Johnston discussed how, because the low-temperature properties are most relevant to the occurrence and mechanism of superconductivity, the classification of the Fe-HTSC as bad or coherent metals should be based on the normal state properties at low temperature [31]. For a conductor with a cylindrical FS, the product $k_F l$ can be written as $k_F l = 0.258 \Delta c/\rho_{ab}$, where $\Delta c$ and $\rho_{ab}$ denote the distance between conducting layers and the in-plane resistivity, respectively [31]. Measurements of the in-plane resistivity in different crystals of the same material $Ba(Fe_{1-x}Co_x)_2As_2$ gave values differing as much as 30% from each other. Anyhow, the value $k_F l \approx 14$ indicates that $Ba(Fe_{1-x}Co_x)_2As_2$ is a coherent metal, and not a bad metal. The estimates of $k_F l \approx 0.5$, which suggested that the Fe-HTSC were bad metals, have been based on the agreement with measurements on polycrystalline samples, now known to be incorrect in light of more recent results collected in single-crystal samples [31].

In addition to the itinerant electron character revealed by Hall effect [82] and nuclear magnetic relaxation experiments [83], further evidence that the Fe-HTSC are coherent metals is provided by the results of quantum oscillations in the magnetization (de Haas van Alphen effect, dHvA) and/or in the resistivity (Shubnikov-de Haas effect, SdH) measured as a function of applied magnetic field. High quality crystals and large mean free paths, i.e. coherent electrons, are necessary requirements for the observation of dHvA and/or SdH oscillations [84,85]. These measurements have been carried out for 1111 LaFePO [84] and 122 compounds such as $SrFe_2As_2$ [86], $BaFe_2As_2$ [87], $CaFe_2P_2$ [88], and $SrFe_2P_2$ [85]. Although the multi-band nature of the Fe-HTSC makes the interpretation of these experiments somewhat difficult, in general there is agreement between the experimental results and the prediction of density functional theory (DFT) calculations of a reconstructed FS arising from a nested-FS-driven SDW. The mass enhancement is found to be rather modest, of order of 1 to 2 times the band structure values calculated with DFT [31,32]. On the contrary, dHvA experiments for $KFe_2As_2$ revealed discrepancies between the calculated and measured FS [89]. The measurements indicated that, depending on the particular band, the carrier mass is strongly enhanced with respect to the respective calculated band mass. These results, ascribed to band narrowing arising from the local Coulomb correlations in the Fe 3d shell, are qualitatively consistent with the large enhancement of the normal state electronic linear heat capacity coefficient reported in ref. [90]. Johnston points out that “despite these sometimes very large carrier mass enhancements, the corresponding conduction carrier conduction must still be coherent since that is required for observation of the dHvA effect.” [31].

Angle resolved photoemission (ARPES) investigations have identified features of the electronic structures which are found to be either in agreement [91,92 93,94], or incompatible with the results of DFT calculations [95,96]. In general, Hubbard bands are not found [91,97], but the $d$-bands appear to be narrower than those obtained from DFT calculations, and a renormalization of ~ 2 is typically required to obtain an agreement between the experimental and the calculated sets of $d$-bands.

The results of x-ray spectroscopy experiments such as core level photoemission (PES), x-ray absorption (XAS) and x-ray emission (XES) spectroscopies are particularly relevant in the context of this discussion since they are expected to elucidate the role of electron correlations. In these experiments, photon absorption causes an electron from a specific core level to either leave the system, as in PES and XES experiments, or to occupy one of the unoccupied levels above the Fermi level ($E_F$), as in the case of XAS. The photo-excited core electron leaves behind a core hole which has the effect of pulling some of the unoccupied states below $E_F$. In this case, in the valence band there appears an effective positive charge that needs to be screened for the system to relax back to a state of minimum energy. When electron correlation effects are at work, different screening channels become available. For example, if a core hole is created in a transition metal with configuration $d^n$, the positive charge in the valence band can be screened either by the electrons in the TM s/$d$-band, or by electrons belonging to bands of the surrounding ligand atoms. The important fact is that different screening channels leave definite signatures in PES, XAS and XES spectra typically in the form of additional satellite peaks.

The Fe spectra obtained from x-ray spectroscopy experiments in 1111 and 122 materials have not displayed additional satellite peaks commonly associated with a localized character of the 3d electrons and indicative of strong on-site Coulomb repulsion as found for example in the Cu 2p core level PES spectra of the cuprates or Fe oxides (cf. Fig. 3). Fe PES and XAS spectra are characterized by lineshapes more akin to those of Fe metal and inter-metallic compounds, and by a high density of states (DOS) at $E_F$, in stark contrast to the case of correlated oxides [55,98,99,100,101,102,103], as shown in Figs. 3 and 4. These results are more in agreement with the existence of weak electronic correlations, with the spectral shapes often being a good match to the occupied/unoccupied electronic DOS determined from standard DFT calculations over a large energy range. For example, Fig. 5 shows a comparison between the total and partial DOS of the valence band extracted from data collected in the 1111 material and DFT calculations. The data indicate substantial overlap between Fe 3*d* and As 4*p* states and between O 2*p* and Ce 5*d* states, in agreement with the calculations [100]. Similar agreement between DFT and spectra is found in XAS measurements of $Ba(Fe_{1-x}Co_x)_2As_2$ [99]. The XAS spectra at both the Fe and As edges are well matched by the unoccupied DOS, indicating the occurrence of weak to moderate electronic correlations, as also suggested by the Fe spectral lineshape, which is similar to that of Fe in compounds with a delocalized character of the 3d states (cf. Fig. 6). The data and the calculations indicate a substantial degree of Fe-As hybridization, as revealed by the strong overlap of Fe $d_{xz+yz}$ states with As $p_{x+y}$ states [99]. This is particularly important since the Fe $d_{xz+yz}$ states contribute substantially to the DOS at $E_F$. Results in accord with DFT predictions of the orbital characters away from $E_F$ are also found [99]. The good agreement between DFT calculations and experimental data indicate that the correlation effects may not play an important role in these compounds on the energy scale probed by x-ray spectroscopy experiments. Interestingly, it appears that the unoccupied states do not exhibit the renormalization effects which affect the occupied electronic bands below $E_F$ as exposed by several ARPES investigations. It should be emphasized, however, that x-ray spectroscopy experiments probe a much wider energy range, extending beyond the energy of the main *d* bands, which may well be renormalized. XAS and XES spectra for different pnictides families have been also directly compared to theoretical calculations that included the presence of the core hole in the absorption and emission processes in full detail, with the results indicating that the Fe-HTSC are weakly correlated materials [55].

It thus appears that that despite some known problems of DFT calculations, such as a strong overestimation of the magnetic tendency of these materials and difficulties describing the interplay between magnetism and Fe-As bonding [104], the description of the electronic structure concerning the orbital occupancies and their relative energies in spectra are not strongly modified by electron correlations. These findings are quite different from what is expected in strongly correlated systems, such as the cuprates, and impose stringent constraints on theories capable of providing a correct description of Fe-HTSC materials. The conclusion is that, from the point of view of x-ray spectroscopy experiments, pnictides Fe-HTSC are weakly, or moderate at most, correlated systems.

Nonetheless, the renormalization effects found in ARPES experiments also indicate that electronic correlations are not negligible. It is thus useful to have quantitative measurements of the level of correlation that exists in these materials. Discussion of the degree of electron correlations in Fe-HTSC is often phrased in terms of the Hubbard *U*. As discussed above, several results values of *U* have been used to describe the Fe-HTSC as everything from weakly to moderately or even strongly correlated systems. In addition to possible conflicts between experimental data as previously discussed, much of the variation of the values of *U* comes from ill-defined use of the term *U* itself, which has quantitative meaning only in the context of the specific model being discussed. It is particularly difficult to find strict correspondence between experimental measurements, in which the orbitals/sites involved cannot always be controlled, and theoretical models in which the orbitals/sites are limited and sharply defined, and vary with the model.

With these limitations in mind, the degree of Coulomb repulsion between holes in the valence band of 1111 and 122 pnictides has recently been probed with PES by measuring core-valence-valence (CVV) Auger transitions [105]. The Auger process is schematically illustrated in Fig. 7. Following the creation of a hole in a deep core level, the system restores a minimum energy configuration by filling the core hole with an electron occupying a higher energy level and promoting a second electron, i.e. the Auger electron, to the continuum. The Auger effect thus leaves the system in a two-hole final state. When the final state consists of both holes in the valence band (VB), the Auger process is labeled as core-valence-valence (CVV). In absence of electron correlations the lineshape of the CVV Auger spectrum resembles the self-convolution of the occupied local DOS, and the spectra are referred to as "band-like" [106]. In this case, if $E_B(V_1)$ and $E_B(V_2)$ denote the binding energy (BE) of two electrons in the VB, and if $E_B(C)$ denotes the binding energy of the core level involved, the kinetic energy (KE) of the Auger electron in a CVV Auger transition is given by KE = $E_B(C) - E_B(V_1) - E_B(V_2)$. When the hole-hole Coulomb repulsion in the Auger final state is not negligible, effects beyond the single particle approximations need to be taken into account, and the KE of CVV Auger electrons can be written as KE = $E_B(C) - E_B(V_1) - E_B(V_2) - \mathcal{U}(V_1,V_2;X)$, where $\mathcal{U}(V_1,V_2;X)$ is the *effective* Coulomb interaction or correlation energy of the two holes in the final state X [107]. Simply put, with respect to the case of band-like spectra, the energy of the CVV Auger electron is lowered by an amount equal to the hole-hole correlation energy, which can be estimated by measuring the KE of CVV Auger spectra. The hole-hole interaction energy is typically written as $\mathcal{U}(V_1,V_2;X) = \mathcal{U}^0(V_1,V_2;X) - R(V_1,V_2;X)$, an expression clarifying that the bare intra-atomic Coulomb energy $\mathcal{U}^0(V_1,V_2;X)$ of interaction between the two holes in the final state is reduced in a solid by a variety of relaxation and screening effects described by $R(V_1,V_2;X)$ [108]. A measure of the electron correlations that exist in these systems is provided by comparing the two-hole final state spectra to the self-convolution of the occupied single-particle, local density of states (SCDOS) obtained from DFT calculations. Fig. 8 shows a direct comparison between the Auger spectra and the SCDOS calculated for Fe orbitals. The difference of the centroids of the two curves provides a measure of the effective Coulomb hole-hole correlation strength. Values between the 122 ($BaFe_2As_2$) and 1111 (CeFeAsO) families and within the 122 family are compared as a function of doping. The data reveal differences between the 1111 and 122 families and even a small variation as a function of the doping $x$ in $Ba(Fe_{1-x}Co_x)_2As_2$ [105]. The values of $\mathcal{U}$ range between $\mathcal{U} \approx 1.9$ eV and $\mathcal{U} \approx 2.5$eV, corresponding to $\mathcal{U}/W \approx 0.3$ and $\mathcal{U}/W \approx 0.4$, respectively, in units of the Fe bandwidth $W$. The magnitude of the shifts and the shape of the Auger lines are markedly different as compared to those measured in strongly correlated systems with $\mathcal{U}/\mathrm{W} \approx 1$ such as cobaltates, cuprates, and manganites [109,110,111,112].

Before commenting on the degrees of Coulomb-type electron correlations in pnictides as exposed by the CVV Auger experiment, it is important to address a couple of remarks. First, and most importantly, the substantial itinerancy in these systems has to be recognized, and it would be incorrect to interpret the results of the experiment as a direct measure of the Hubbard "*U*". As stated above, "*U*", is a theoretical construct that is model-dependent, i.e. the term "*U*" has quantitative meaning only in the context of the specific model being discussed. In theoretical models the orbitals/sites are limited and sharply defined, and vary with the model. On the contrary, the CVV experiment cannot distinguish different Fe orbitals, and the value of the effective hole-hole Coulomb interaction $\mathcal{U}$ is thus averaged within the unfilled 3s-3d shell of the Fe atom. This is the reason why the measured effective $\mathcal{U}$ is denoted with a different symbol than the Hubbard "*U*".

Consequently, a compelling issue is the identification of the model, or basis set, that best corresponds to the orbitals probed by the experiment. Providing a specific answer to this question is problematic because the calculated centroid of the SCDOS depends on how the local DOS is calculated. An ideal calculation (and corresponding model) would take into account precisely the set of orbitals occupied by the two holes involved in the final CVV state, but a full knowledge of how to describe these orbitals is unavailable. In his original formulation Sawatzky reasonably assumed Wannier functions localized to the relevant atom [113]. Nonetheless, any method based on Wannier functions is non-unique for a multi-band system, as the "*U*" value corresponding to a Wannier function-based model depends strongly on how many Wannier orbitals are included [114,115]. In ref. [105], the authors used a projection method to select out contributions from states within a sphere around the Fe atom that have *d*-like symmetry. The resulting $\mathcal{U}$ should therefore be conceived of as the difference between the measured Auger spectra and the SCDOS resulting from *specifically* these selected orbitals. This particular choice of orbitals for the calculation of the SCDOS however is highly restrictive, since including any other states would shift the centroid of the calculated SCDOS to higher values on the two-hole scale, therefore reducing the value of $\mathcal{U}$.

As pointed out by the authors, although the value of $\mathcal{U}$ should not be understood to provide a quantitative value for any specific Hubbard or Hubbard-based model requiring a "*U*", it provides an upper bound for the Hubbard *U* [105]. In light of these results and relative considerations, the interaction between holes in the VB of the studied pnictide compounds is highly screened, suggesting that from the point of view of the strength of the on-site Coulomb repulsion these systems can be considered as weakly or moderate at most, and certainly not strongly correlated. Although the reported values of $\mathcal{U}$ are not suitable for direct use in model calculations of the Hubbard type, they provide constraints on theoretical descriptions of phase diagrams that vary with the quantity *U* or *U/W* by providing upper bounds to *U* or *U/W*.

On the other hand, certainly electron correlations cannot be dismissed, as suggested by the renormalization of ~ 2 that is necessary to apply to the electron bands revealed by ARPES in order to have a satisfactory correspondence with DFT band structure calculations. Although it has been argued here that the values of Coulomb type electron correlations are moderate at most, the possibility that such correlations exhibit a momentum dependence cannot be dismissed. If this were the case, this occurrence would not be identified in integrated x-ray spectroscopy experiments, which on the contrary provide an estimate of the electron-electron interactions mediated in momentum space and generally also among different orbitals. It has been suggested that intermediate values of the Hubbard *U* need to be explored in more detail as they might reveal paradigms not previously explored [34]. In addition to the above mentioned Coulomb-type correlations, it is necessary to consider other type of electron correlations, such as magnetic correlations that couple to the *d*-electrons and can lead to renormalizations and shifts in spectral weight within the d-bands, with coupling through the exchange correlation term (Hund's rule), as discussed for example in Refs. [116,60,117,118]. Hund's magnetic correlations, or Hund's coupling, refers to the intra-atomic exchange interaction that results in lowering the energy of electrons with parallel spins in different orbitals, as opposed to having the electrons with antiparallel spins in the same orbital as in Hubbard-type models. The relevance of magnetic correlations for explaining the magnetism in pnictides is described in the next section.

## 3. THE MAGNETIC MOMENT IN PNICTIDES

### 3.1 The magnitude of the magnetic moment in pnictides

The possible crucial role played by magnetic correlations in the physics of Fe-HTSC was mentioned in the Introduction. Besides the proximity of the superconducting and magnetically ordered states, which suggests that the superconducting pairing mechanism may be related to the coexistent magnetism in the phase diagram, the importance of the microscopic origin of magnetism in Fe-HTSC is intimately tied to the electronic structure and the degree of electron correlations existing in these materials.

Shortly after their discovery, the localized versus itinerant nature of magnetism in Fe-HTSC was a matter of controversy. Experiments including x-ray spectroscopy [98,99,55], De Haas-van Alphen [84,85], Hall effect [82], and nuclear magnetic relaxation [119] revealed an itinerant electron character. On the contrary, neutron scattering experiments seemed to be well described by a local picture, with local magnetic moments on the Fe sites [120,121]. The occurrence of localized and itinerant characters of electron spins suggested by different experiments has been at the beginning often interpreted and presented as an axiomatic dilemma between two extremely different theoretical viewpoints [122]: The first one described the electrons as localized, while the other viewpoint identified the Fe-HTSC as itinerant magnets. In real materials, any proper description of magnetism falls on a spectrum having as limits a fully localized picture on one side, and a fully itinerant picture on the other side. Mott-Hubbard-type arguments tend to fall on the localized side of the spectrum, while arguments based on Fermi surface nesting falls on the fully itinerant side of the spectrum. Immediately after the discovery of Fe-HTSC, the discussions on the nature of magnetism centered around the two limits of the spectrum, that we now describe.

Within the localized viewpoint, the magnetic moments are formed due to the intra-atomic exchange interaction of the electron belonging to the outermost unfilled atomic shells. In the parent compounds of Fe-HTSC, simple valence counting arguments indicate that Fe is in a formal oxidation state $Fe^{2+}$, with a $d^6$ electronic configuration. Consider the single $Fe^{2+}$ ion first. In this case, the magnitude of the magnetic moment depends on the magnitude of the total angular momentum $\vec{J}$ of the unfilled atomic shell. The expression for the momentum $\vec{J}$ depends on the different ways in which the orbital and spin angular momenta of the electrons can be coupled together. The two most common schemes are known as *LS*-coupling and *jj*-coupling, which correspond to the two extreme limits of the strength of the spin-orbit interaction. For light atoms, the spin-orbit interaction is usually weak, and so it can be treated as a perturbation. This is the range of validity of the *LS*-coupling scheme, according to which one considers first the separate coupling of the orbital $\vec{l}_i$ and spin $\vec{s}_i$ angular momenta of each of the *N* electrons so as to yield the total angular momenta $\vec{L} = \sum_{i=1}^{N} \vec{l}_i$ and $\vec{S} = \sum_{i=1}^{N} \vec{s}_i$ . The spin orbit interaction between $\vec{L}$ and $\vec{S}$ is then considered perturbatively at this stage, resulting in the formation of the total angular momentum $\vec{J} = \vec{L} + \vec{S}$ . In the opposite limit where the spin orbit interaction becomes dominant, as in the case of high Z-atoms, one considers first the coupling between the orbital $\vec{l}_i$ and the spin $\vec{s}_i$ angular momentum for each electron so as to yield a total angular momentum $\vec{j}_i = \vec{l}_i + \vec{s}_i$ for each electron. The total angular momentum of the shell is then obtained by composing the individual angular momenta $\vec{j}_i$ as $\vec{J} = \sum_{i=1}^{N} \vec{j}_i$ .

Since in 3d systems the spin-orbit interaction is not predominant, for the $Fe^{2+}$ ion the total angular momentum $\vec{J}$ is determined in the *LS*-coupling scheme. The magnetic moment is given by $\mu = \mu_B g J$, where $g$ denotes the Lande' factor $g = 1 + \frac{J(J+1) + S(S+1) - L(L+1)}{2J(J+1)}$ . For transition metals like Fe, the angular momentum is quenched, i.e. $L = 0$, and thus $J = S$ and $g = 2$. The moment of the $Fe^{2+}$ ion is thus given by the spin $S$ only as $\mu = 2\mu_B S$, and for this reason it is referred to as a spin moment (*SM*). As stated by the first Hund's rule, the ground-state energy corresponds to the maximum value of $S$. The *SM* $\mu = 2\mu_B S$ is thus given by Hund's rule, which is a consequence of the fact that the dominant interaction is the intra-atomic exchange.

This situation applies to the isolated $Fe^{2+}$ ion. When the $Fe^{2+}$ ion is surrounded by a configuration of ligand atoms, the spherical symmetry proper to the atomic case is broken, and the five-fold degeneracy of the $d$ levels is lifted, resulting in the $d$-manifold levels being split into a twofold degenerate and a threefold degenerate group of levels denoted as $e_g$ and $t_{2g}$, respectively. The energy difference between the $e_g$ and $t_{2g}$ levels is known as the crystal field, and is denoted as *10Dq*. In this case, the contribution to the *SM* depends on the competition between the exchange energy and the crystal field *10Dq*. In a tetrahedral environment, in which the $e_g$ levels have lower energy than the $t_{2g}$ ones, two spin states are expected: The low spin state ($e_g^4 t_{2g}^2$), and the high spin state ($e_g^3 t_{2g}^3$), with spin values of $S = 1$ and $S = 2$, and local *SM* with values 2 $\mu_B$ and 4 $\mu_B$, respectively. The inter-atomic exchange, i.e. the exchange interaction with the neighboring atoms, is smaller than the intra-atomic exchange, but it is important since it leads to the magnetic ordering. According to the localized viewpoint, anti-ferromagnetism stems from ordering of the local SM via short-range super-exchange [123].

The other viewpoint identified the Fe-HTSC as itinerant magnets. In the framework of itinerant magnetism, the magnetic moments are not given by the angular momentum, but they originate from the delocalized itinerant valence electrons. In this case the driving interaction is the inter-atomic exchange among the conduction electrons. According to this viewpoint, the Fe-HTSC are considered itinerant weakly correlated metallic systems which become magnetic via nesting between the hole and the electron pockets in the FS [124].

These two viewpoints are on the opposite sides of the spectrum of possible approaches for the description of magnetism in Fe-HTSC. In fact, Haule and Kotliar have emphasized the importance of both correlations and metallic behavior since the beginning of the field [60,116]. The same authors have proposed a scenario featuring a crossover from coherent itinerant magnetism to incoherent local moment magnetism occurring at a temperature T* [116]. DFT is somewhere on the spectrum, and where precisely it falls depends on the type of functional used and different approximations. Theoretically estimated value of the *SM* are found to be $\approx 2\mu_B$ as calculated with DFT and slightly larger ($\approx 2.4\ \mu_B$) according to DMFT combined with DFT [31].

Until 2010 the majority of work has been carried out in the 1111, 122 and 111 families. As discussed above, most experimental data have suggested an itinerant electron character which did not seem to support the possible existence of local SM. Measurements in the ordered phases have found small values of the SM and a significant considerable variation of the SM values within different families. The SM were found to be considerably smaller than those suggested by theoretical predictions [125].

The small values of the SM and the variability of its size seemed to be inadequate for descriptions based on local moment models, shaping within the majority of the community the conclusion that the magnetic order does not stem from exchange interactions between local magnetic moments with fixed magnitude. This conclusion was also indicated by the fact that the detection of SM in the paramagnetic phase, an observation which would indicate the validity of local moment pictures, has remained elusive for quite some time.

This perspective of the nature of magnetism and the character of electron correlation in Fe-HTSC is slowly changing in the last few years primarily in light of two factors. One factor is the discovery of new families of compounds such as the "11" chalcogenides and the $A_xFe_{2-y}Se_2$ alkali metal iron selenides, in which strong electron correlations effects and large local spin moments have been detected. The other factor is the provision of new data providing estimates of the spin moment based on different type of experiments. It is therefore important to report on these data.

Consider first the 11 family. The values of the SM in the FeTe 11 compounds reach $\approx 2.5\ \mu_B$ in the parent compounds [31,34]. These values in the 11 compounds are significantly larger than those in the pnictides, suggesting important differences with the latter materials. As discussed above, a local moment picture for the magnetism appears to be appropriate for the 11 compounds, in agreement with the presence of stronger electron correlations [34,70]. Fe SM have also been detected with x-ray emission spectroscopy (XES) in the paramagnetic (PM) phase of a number of chalcogenides [126]. The SM values are consistent with the values of static moments measured experimentally, in agreement with a local moment picture. The occurrence of local moments found in chalcogenides is consistent with the presence of stronger electron correlations [34,70].

The picture for the pnictides is less clear and needs to be examined in detail. Until 2010, experimental values were found to be small and to vary considerably, as indicated for example by the values of the ordered SM per Fe atom at low temperature reported in Table 10 of reference [31]. The SM per Fe atom measured in ordered magnetic phase ranges from 0.25 $\mu_B$ to 1 $\mu_B$, i.e. with an upper value for the spin S = ½, definitively smaller than the values of 2 $\mu_B$ or 4 $\mu_B$ corresponding to spins S = 1 or S = 2, respectively. More specifically, the ordered Fe SM ranges from 0.25 for NdFeAsO (Ref. [127]) to ≈ 1 $\mu_B$ in $SrFe_2As_2$ [128] for the 1111 and 122 pnictides, and ≈ 0.09 for NaFeAs. Inelastic magnetic neutron scattering (INS) have provided direct estimates of lower limits of the effective moments in the paramagnetic state of $CaFe_2As_2$ and $Ba(Fe_{1.935}Co_{0.065})_2As_2$, but the values of the moments were found to be quite small, i.e. 0.31 $\mu_B$/Fe and 0.47 $\mu_B$/Fe, respectively [129,130]. These results have suggested instead that the mechanism responsible for the AFM order is to be found in the SDW arising from itinerant electrons, specifically from nesting of the hole and electron pockets of the FS. The FS topology compatible with this picture has been observed in ARPES experiments [131,132], consistent with the results of quantum oscillation measurements [87,133].

Interestingly, while in general DFT underestimates the magnitude of the ordered SM, in the pnictides the opposite happens, with an estimated value ≈ 2$\mu_B$ [125]. Other lines of research have thus proposed that the occurrence of fast fluctuations of the SM could provide a rationale for the theoretical overestimation of the ordered moments [134,135,136]. The small values of the SM observed by INS experiments in the paramagnetic state, however, seemed not to support this picture. On the theoretical front, Haule and Kotliar have proposed a scenario featuring a crossover from coherent itinerant magnetism to incoherent local moment magnetism occurring at a temperature T* [116], but the predicted temperature evolution of the magnetic susceptibility, from Pauli-like to Curie-Weiss-like, has been found not to be in agreement with experiments [31,137]. Other theoretical studies which specifically addressed the role of magnetic frustration and fluctuations have argued against a local moment picture [138, 139]. All of these observations have understandably shaped the general belief that in the pnictides itinerant SDW and FS nesting are usually providing valid mechanisms for describing the magnetic states. The underlying rationale is the occurrence of an itinerant electron character, which is incompatible with the occurrence of strong local Mott-Hubbard type correlations that for example characterize the cuprates. Although this is the general belief, it is important to stress that, in fact, even within the itinerant picture some authors have proposed arguments against considering FS nesting as a valid mechanism for the description of the magnetic states [122].

Since 2009, several works have proposed that the Fe *d* electrons have both localized and itinerant characters. On the experimental front, neutron scattering studies have suggested the importance of both localized and itinerant Fe *d* electrons [140, 141], and similar conclusions have been inferred from optical spectroscopy [142,143,144]. Theoretical works have indicated the necessity to attain descriptions beyond a fully localized or completely itinerant perspective [122, 136,141,145,146,147,148,149,150,151,152,153,154,, 155156,157,158,159], pointing out in particular the inadequacy of a pure FS nesting picture.

The provision of relatively recent x-ray spectroscopy data indicates that in the pnictides the microscopic origin of magnetism appears to be more enigmatic than in the 11 family. Fe local SM have been detected with x-ray emission spectroscopy (XES) in the paramagnetic phase of different pnictide compounds, namely LiFeAs (0.9 $\mu_B$/Fe), PrFeAsO (1.3 $\mu_B$/Fe), $BaFe_2As_2$, (1.3 $\mu_B$/Fe), $Ba(Fe_{0.915}Co_{0.085})_2As_2$ (1.1 $\mu_B$/Fe) (Fig. 9). These values of the SM are slightly larger, but consistent, with the ordered moment measured in the 122 compounds, but definitely larger than those found in 111 and 1111 compounds [126]. The values of the SM measured with XES in the paramagnetic phase are larger than the values obtained with INS in the paramagnetic state of $CaFe_2As_2$ (0.31 $\mu_B$/Fe) [129], $Ba(Fe_{1.935}Co_{0.065})_2As_2$ (0.47 $\mu_B$/Fe) [130] and, although in better agreement, $BaFe_{1.9}Ni_{0.1}As_2$ (≈ 1 $\mu_B$/Fe) [159]. Although these values for the SM provided with XES and INS in $BaFe_{1.9}Ni_{0.1}As_2$ are still smaller than the values calculated with DFT (≈ 2$\mu_B$/Fe) or DMFT (2.4 $\mu_B$/Fe), it is now possible to claim that the observation of non-negligible local SM in the paramagnetic phase of pnictide compounds has been established. This observation is important, since the presence of local moments in the paramagnetic phase is incompatible with theories relying exclusively on FS instabilities, e.g.

SDW, since in these frameworks the processes of moment formation and long range ordering should occur concomitantly at $T_{N}$.

More recently, a direct and element-specific measurement of the local Fe spin moment has also been provided by analyzing the Fe 3*s* core level photoemission (PES) spectra in the parent and optimally doped $CeFeAsO_{1-x}F_x$ (x = 0, 0.11) and $Sr(Fe_{1-x}Co_x)_2As_2$ (x = 0, 0.10) pnictides [160]. The magnitude of the SM is found to be 1.3 $\mu_B$ in CeFeAsO, and 2.1 $\mu_B$ in $SrFe_2As_2$, the latter value being in much better agreement with the results of theoretical predictions.

The detection of the largest values to date of the SM in pnictides is thus provided by soft x-ray spectroscopy measurements such as PES and XES. The detection of SM in XES and PES experiments is based on the same mechanism that, given the importance of the observation of local SM for the physics of the pnictides, we now describe. This mechanism is based on the multiplet splitting (M-SP) effect following the creation of a core hole due to photon absorption. The M-SP effect can occur only in systems in which the outer subshell(s) are partially occupied with a non-vanishing spin $S_V$. The M-SP effect arises from the coupling of the core electron left behind upon photoelectron emission with the net spin $S_V$ in the unfilled shells of the emitter atom. Fig. 10 shows a schematic layout of the energy levels involved in PES experiment in the case a hole is initially created in an "*s*" core level. The analysis of M-SP effects is considerably simplified for "*s"* core level spectra since in this case the core hole has zero angular momentum, thus limiting the number of possible final states [161,162]. Upon photoemission of one electron from the inner 3s core level, two final states for the ion are possible, corresponding to the configurations in which the remaining core *s* electron is either parallel or anti-parallel to the net spin $S_V$ in the unfilled *3d-4s/p* shell of the TM atom. The exchange energy of the state with parallel spins is lower than that with anti-parallel spins, and consequently the energy difference between these two states is revealed as the presence of a double peak in PES experiments.

The double peak structure provides the possibility of extracting the net spin $S_V$ in the outermost shell(s) of the TM atom. In the most simple interpretation, which is valid for ionic systems, the multiplet energy separation $\Delta E_{3s}$ depends on the net spin $S_V$ of the emitter atom via $\Delta E_{3s} = (2S_v + 1)J^{eff}_{3s\text{-}3d/4s}$, a result known as Van Vleck theorem. $J^{eff}_{3s\text{-}3d/4s}$ denotes the effective exchange integral between the 3s and the 3d/4s shells after allowing for final-state intra-shell correlation effects [162]. Although the analysis of 3s core levels has usually been carried out for ionic compounds, multiplet exchange splittings are also detectable in metallic systems. Previous work on metallic systems such as Mn and Co has shown that, although Van Vleck's Theorem is insufficient to describe properly the itinerant nature of the electrons, $\Delta E_{3s}$ is found to scale linearly with ($2S_V$ + 1), indicating that the *3s-3d/4s* exchange interaction is the dominant contribution of the lineshape of the 3*s* core level spectra in itinerant systems [163,164,165,160]. For the Fe-HTSC, the magnitude of the net spin $S_V$, and thus the SM $2S_V$ are obtained from the splitting $\Delta E_{3s}$ with a procedure described elsewhere [160]. As far as the XES measurements are concerned, the determination of the local moment hinges on the same mechanism as in PES. In XES spectra following excitation of the Fe K-shell, the Fe 1s core hole is filled with Fe 3p electrons, as shown in Fig. 11. The XES spectrum, which consists of the measurement of the photons emitted in the 3p → 1s dipole-allowed transition, exhibits a doublet, corresponding to the two possible ways the electrons left in the 3p core level couple with the net spin $S_V$ of the outermost shell. The estimation of the SM is carried out by means of a calibration procedure described in ref. [126].

A few observations are in order. Multiplet splitting effects occur exclusively in atoms with the outer subshell(s) partially occupied with a non-vanishing net spin $S_v$. Therefore, the XES and PES results indicate that the electronic configuration on the Fe site is never found to be in a spin state with $S_v = 0$. Since the effect underpinning the detection of SM in XES and PES is the exchange interaction between the net spin in the outermost shell(s) and the core hole, the detection of large SM in the paramagnetic phase by XES [126], and in the paramagnetic, anti-ferromagnetic and superconducting phases with PES [160], are indicative of the occurrence of ubiquitous strong Hund's magnetic correlations. More specifically, the large values of the Fe SM indicate the occurrence of a rather strong on-site Hund coupling $J_H$ that fosters the electrons in the Fe *3d/4s* shells to align parallel to each other, as already suggested by theoretical investigations [148,149,154].

The occurrence of fast fluctuations of the SM were initially dismissed as a rationale for the theoretical overestimation of the ordered SM in light of the low values of the SM in the paramagnetic state provided by the first neutron experiments [31]. On the contrary, the role of fast fluctuations in the SM needs to be reconsidered in light of the more recent INS, XES and PES results. Hansmann and coworkers have discussed the necessity of carrying out direct measurements of the SM with fast probes, since the fluctuations of the SM are predicted to occur over fast timescales comparable to the electron dynamic ($10^{-15}$ s), which are too fast to be measured by ordinary magnetic probes such as nuclear magnetic resonance (NMR), muon spin relaxation ($\mu$SR) and Mössbauer spectroscopy [136]. In fact, there exists a correlation between the measured values of the SM and the experimental technique used. The values for the local SM obtained with INS, XES and PES are significantly larger than the ordered SM detected in the AFM phase by neutron diffraction, NMR, $\mu$SR and Mössbauer spectroscopy. The processes involved in x-ray spectroscopic and INS measurements occur on sub-picosecond time scales, much faster than the $10^{-8}$ s - $10^{-6}$ s time scales typical of conventional magnetic measurements such as NMR, $\mu$SR and Mössbauer, indicating that the discriminating factor involved in the determination of the magnitude of the SM is the time scale of the measurement. This is also true for the fast measurements. That this is the case is also confirmed by an inspection of the INS data collected in the paramagnetic phase. Dynamical information can be obtained in INS experiments from integrating the dynamical spin structure factor S($\vec{Q}$,ω) over energy ω and momentum transfer $\vec{Q}$ [31,166]. The function S($\vec{Q}$,ω) obeys the sum rule $\int_{-\infty}^{\infty} d\omega \int_{BZ} d\vec{Q}\, S(\vec{Q},\omega) \propto$ S(S+1), that is, when integrated over the whole Brillouin zone and over the whole frequency range, the function S($\vec{Q}$,ω) provides the value of the square of the effective instantaneous SM. The function S($\vec{Q}$,ω), which is the Fourier transform of the spin-spin correlation function, can be related to the imaginary part of a generalized spin susceptibility, i.e. S($\vec{Q}$,ω) $\propto$ χ"($\vec{Q}$,ω) [166]. By integrating χ"($\vec{Q}$,ω) over $\vec{Q}$ one obtains the local susceptibility χ"(ω). Integrating χ"(ω) over a large frequency range allows the determination of the instantaneous SM, which thus corresponds to the short time limit of the spin susceptibility $\chi''(t = 0)$. Conversely, an integration of χ"(ω) over a limited frequency range yields the static spin susceptibility $\chi''(t \to \infty)$, which describes the response of the system for long times, and corresponds to a screened SM. As correctly pointed out by Johnston, INS data provide lower limits of the effective moments since the energy integration window in χ''(ω) is finite [31]. Indeed, the different values of the SM found in 122 compounds such as $CaFe_2As_2$ (0.31 $\mu_B$/Fe), $Ba(Fe_{1.935}Co_{0.065})_2As_2$ (0.47 $\mu_B$/Fe) and $BaFe_{1.9}Ni_{0.1}As_2$ (≈ 1 $\mu_B$/Fe) correspond to different energy integration windows: 80 meV for $CaFe_2As_2$, 100 meV for $Ba(Fe_{1.935}Co_{0.065})_2As_2$ and 350 meV $BaFe_{1.9}Ni_{0.1}As_2$. The fact that larger integration windows correspond to larger values of the SM is another indication of the importance of the time scale in measurements of the SM, since larger energy windows correspond to shorter time limits of the spin susceptibility.

Two important facts in comparing INS and PES/XES data need to be kept in mind: First, INS is a fast technique ($10^{-14}$ s), but not quite as fast as PES and XES ($10^{-16}$ s). Second, in order to extract the square of the instantaneous moment S(S+1), the spin structure factor S($\vec{Q}$,ω) must be integrated over the full range of momenta and energy, and such integrations are sometimes challenging to perform [166]. For instance, the scattering can be rather broad in momentum and energy, posing the difficult task of separating the spectral weight due to actual scattering from the background. The loss of meaningful spectral weight due to this occurrence could result in underestimating the value of the SM.

An obvious question at this point is why there is not agreement between the magnitude of the SM determined from XES and PES measurements, given that XES is a fast probe comparable to PES. It has been proposed that itinerant electrons are not properly counted in the XES detection of the Fe SM due to the local nature of the Fe 1*s* core-hole potential [126]. A key quantity which the SM is proportional to is the exchange integral between the wavefunctions of the valence electrons and the core hole, which translates in the degree of their spatial overlap. In the XES experiments reported in ref. [126], a core hole is created in the 1s core level. The overlap of the 1*s* core level and 3*p* valence level wavefuntions is not very significant, thus providing an explanation as to why XES experiments are not able to detect valence itinerant electrons contributing to the SM. In PES experiments, the core hole is typically created in the 3*s* core level, i.e. in the same shells as the itinerant Fe 3*d* levels. The larger overlap of the 3*s* and 3*d* wavefunctions suggests that PES experiments are able to detect more sensitively valence itinerant electrons contributing to the SM than XES experiments. The results of XES measurements thus provide lower limits of the magnitude of the SM.

All of the data presented so far allow one to draw an important conclusion, that is, *the magnitude of the SM is found to increase when the latter is probed with measurements performed on faster time scales*. The rapid time scales of the PES process allowed the detection of large local spin moments fluctuating on a $10^{-16}$ - $10^{-15}$ s time scale in the paramagnetic, anti-ferromagnetic and superconducting phases, indicative of the occurrence of ubiquitous strong Hund's magnetic correlations. Works based on DMFT have predicted the presence of strong on-site exchange interactions (Hund's coupling) which, via spin-fluctuations that couple to the *d*-electrons, can provide a rationale for the occurrence of large local moments and renormalizations and shifts in spectral weight within the Fe *d*-bands [116,60]. On the other hand, in general the presence of local moments is often interpreted as a sign of occurrence of strong Coulomb-type electron correlations, while systems hosting weak electron correlations are typically considered itinerant magnets. This distinction is far from being clear cut, as observed by Johnston [31]. Real materials host a continuum range of degrees of electron correlations, and the nature of magnetism is found to range between the local and completely itinerant extreme limits. This has actually been known for a long time, a result emerging from the Rhodes-Wohlfarth plot [167], here shown in Fig. 12.

The Rhodes-Wohlfarth plot is a phenomenological curve which provides information of possible different mechanisms responsible for magnetic order. The plot consists in the dependence of the ratio $q_c/q_s$ on the Curie temperature $T_{Curie}$ for several ferromagnetic compounds. Here $q_s$ denotes the total magnetization per atom in the ordered state, i.e. below $T_{Curie}$, and it is referred to as the magnetic carrier per atom. It coincides with *gJ* if the moment is localized, i.e. $\mu = \mu_B gJ$. The quantity $q_c$ is still defined as a magnetic carrier, i.e. $q_c = gJ$, but calculated from the expression of the Curie constant, and thus is representative for $T > T_{Curie}$. The ratio $q_c/q_s$ describes the degree to which a system exhibits local or itinerant magnetism. If a systems hosts local moments, the magnitudes of the latter are not expected to change below and above $T_{Curie}$, and thus $q_c/q_s \sim 1$. On the contrary, for itinerant magnetism $q_s$ becomes vanishingly small for systems with low $T_{Curie}$, and thus $q_c/q_s \approx 1$. As shown in Fig. 12, the distribution of the data points suggests that local and itinerant magnetism are two extreme limits that are hardly reached in real materials.

The observation of localized moments and itinerant electrons in Fe-HTSC poses the theoretical challenge of reconciling the localized- and itinerant-electron models for a magnetic, metallic system. This is strongly reminiscent of a problem that faced the scientific community in the late 70', namely the problem of magnetism in metallic Fe. The value of $q_c/q_s = 1$ on the Rhodes-Wohlfarth plot indicates that Fe is a local moment system exhibiting Curie-Weiss behavior, although Fe is certainly a metal [168]. The specific case of the magnetism in metallic Fe will be now described below. The pedagogical value of this example lies in the importance of paying particular attention to the electron dynamics in understanding the presence of local SM in systems hosting itinerant electrons. Considerations on electron dynamics offer some insights for a description of magnetism in itinerant systems and provide a rationale for the presence of both electron itinerancy and local SM exposed by the experiments.

### 3.2 The magnetism in metallic Iron.

Metallic Fe exhibits some behaviors interpretable in terms of band- theory (itinerant-electron) models, others in terms of a localized-electrons model. Revisiting the fundamental steps taken towards the solution of this problem will prove to be very insightful for the description of the Fe-HTSC and, more generally, magnetic systems hosting itinerant electrons.

The Stoner-Wohlfarth model (SWM), also referred to as band magnetism, has been the cornerstone of itinerant magnetism [ 169 ]. The spontaneous magnetization is given by the different occupation of the split up- and down-spin electron bands. Band structure calculations for transition metals account very well for the groundstate properties (i.e. at T = 0 K), with the exchange splitting $\Delta_{EX}$ given by the energy difference between up-and down-spin electron *d* bands [170]. Yet, the SWM fails miserably in explaining the physical properties of the archetypal metallic ferromagnet, metallic Fe. The inadequacy of itinerant models in describing metallic Fe is at best illustrated by considering the problems of calculating the Curie temperature $T_{Curie}$ and explaining the existence of a Curie-Weiss moment above $T_{Curie}$. Calculations of the Fe band structure account very well for the groundstate properties, with a value of the exchange splitting $\Delta_{EX} \approx$ 2 eV. Nonetheless, band theory fails when one attempts to estimate the Curie temperature from $k_B T_{Curie} \approx \Delta_{EX}$, since one would get a value for $T_{Curie}$ much larger than the experimental value $T_{Curie} \approx 1000$ K. The theory also predicts no moments and no Curie-Weiss law above $T_{Curie}$, in sharp contrast to experimental results [168].

It is now well understood that the reason for this is that the only excited states contemplated in band theory are the Stoner excitations between the spin polarized electron bands. Conventional band theories based on DFT fail to describe the total magnetization correctly since the magnetization in each unit cell point in the same direction. Within this framework, the total magnetization may vanish only if the exchange splitting, and thus the local moment, vanishes.

Giving the experimental value of $T_{Curie}$, which is much lower than the value predicted by band theory, it is expected that there exist excitations with much lower energy than the Stoner excitations. All modern theories for metallic Fe should contemplate the fact that the driving factor for the magnetic-to-paramagnetic state transition is the fluctuations of the spins direction, whereas magnitude fluctuations of the local spin density are of minor importance due to the extremely high energy cost [171,172,173,174,175]. The magnetization direction must be allowed to vary from unit cell to unit cell. At $T_{Curie}$, the magnetization thus vanishes not because the magnetic moments vanish (as in the case of the Stoner excitations), but because the latter are no longer oriented parallel to each other, rather pointing in every direction with equal occurrence. These fluctuations in directions constitute a set of excited states with much lower energy than the Stoner excitations. This type of excitations, quite reminiscent of those described in the localized models, must be properly described in a framework where electrons are itinerant, and this fact poses a real challenge.

Hubbard and Hasegawa were among the first to propose an amalgam of localized and itinerant models when studying the magnetism in metallic Fe [173,174,175]. In their proposed theory, the electrons are described as itinerant, but they are influenced in their motion by "exchange fields" configurations localized at atoms which correspond very roughly to the spin configurations of the localized models. Because of the short range nature of the exchange interaction, Hubbard proposed that the exchange fields are entities essentially proportional to the magnetic moments of the atoms. The motion of the itinerant electrons and the configurations of the exchange fields are influenced reciprocally in a self-consistent fashion. The low energy excited states are such that the exchange fields vary from atom to atom, and thus not properly describable in ordinary band theories [173,174,175].

It is then natural to ask whether standard band structure theory can describe these low energy excited states. The formal foundation of spin-polarized band theory is the spin-density functional description of the inhomogeneous electron liquid in the periodic electrostatic field of the nuclei and the local spin-density (LSD) approximation for the exchange-correlation energy. The exchange correlation potential commonly used in DFT corresponds to Hubbard's concept of exchange fields. The problem of standard DFT calculations in describing these low energy excited states is that DFT is a mean field theory. The mean field approximation suppresses entirely the transverse fluctuations in space necessary to properly understand magnetism in itinerant systems. In principle, a more sophisticated, non-local approximation for the exchange correlation potential in place of the LSD scheme may overcome this well-known inadequacy of the spin-density functional approach.

For the particular problem of Fe metal, it has been possible to cast the idea of exchange fields proposed by Hubbard in a form most suitable for first-principles calculations [176]. Band theory based on the LSD approximation has been generalized to finite temperatures by treating the magnetic fluctuations in the molecular field approximation familiar in the context of spin-only Hamiltonians. The success of this approach stems from paying particular attention to the electron dynamics in itinerant systems [176].

In the insulating transition metal oxides and in rare earth metals, localized magnetic moments form from well localized electronic wavefunctions not participating in the Fermi surface. In this case, the magnetism can be discussed concentrating on the magnetic degrees of freedom alone, typically described by spin Hamiltonians (such as the Heisenberg Hamiltonian). The case of itinerant systems is more complicated. The magnetism in systems like Fe originates from itinerant *d*-electrons which also happen to participate in the Fermi-surface, so that the separation between magnetic and translational degrees of freedom does not occur. For a discussion of the magnetic properties, one has to deal with the full many-body problem, this complication probably having hampered the development of an understanding of magnetism in itinerant systems.

A key characteristic of itinerant systems is that the amplitude of the magnetic moment is not constant but exhibit very fast *quantum fluctuations*. Itinerant electrons have wavefunctions which are phase-coherent over large distances, with the result that the electron density, and as a consequence the spin density, are not described by sharp quantum numbers. If $W$ denotes the bandwidth, which is typically a few eV, itinerant systems are characterized by the presence of a fundamental time scale of the order of $\tau_Q \approx h/W$, i.e. the hopping time of electrons from site to site, which produce very fast quantum fluctuations. The existence of this typical fast time constant $\tau_Q \approx 10^{-15}$ s is characteristic for itinerant systems and has no equivalence in localized magnetism. The expression "very fast quantum fluctuations" indicate that these fluctuations manifest directly in fast experiments with a short time constant of the order of $\tau_Q$, and thus involving large energy transfer. A "slow" measurement with a typical time constant larger than $\tau_Q$ (which means typical energy transfers much smaller than $W$) will average over these quantum fluctuations and detect the average moment per atom [172].

These quantum-averaged moments may fluctuate slowly in time if the system is in some excited configuration. The spin wave configuration corresponds to a slow wave-like precession of the atomic moments averaged over the fast quantum fluctuations. Given that the typical spin wave energy is $W_{SW} \approx 50$ meV, the time constant $\tau_{SW}$ associated with spin wave motion is $\tau_{SW} \approx h/W_{SW} \approx 10^{-13}$ s, much slower than the fast $\tau_Q$. This clear separation of timescales makes a visualization of the process of moment formation possible. On a time scale $\tau$ long compared to $\tau_Q$, but short compared to $\tau_{SW}$, electrons arrive at and leave a site with sufficient correlation between their spin orientations so as to yield a non-vanishing magnetic moment. Subsequently, on the time scale comparable to the spin wave motion $\tau_{SW}$, the moments exhibit a slow motion in which they can change their orientation, as in thermal fluctuations described so well by spin-Hamiltonians. This is the moment that is observed above $T_{Curie}$ as a Curie-Weiss behavior.

The occurrence of different time scales and the distinct separation between fast and slow electron motion constitute the rationale for tackling the problem of magnetism in itinerant systems from first principle, as demonstrated for the case of Fe metal [176]. The time scale $\tau$, long compared to $\tau_Q$, but short compared to $\tau_{SW}$, constitutes a time window in which the state of the system can be described in terms of orientational configurations specified by assigning a set of directions for the magnetization in each unit cell. For a fixed orientational configuration the problem is treated in the local spin density approximation, as it is customary in spin-polarized band structure calculations for T = 0. The theory reduces to the conventional LSD calculations at T = 0 where all the moments are lined up, with no adjustable parameters other than the atomic number and the lattice constants. The determination of the changes in the orientational configuration of the moments allows the description of the slow motion of the moments. The time evolution of the system configurations is described based on the assumption that the system is ergodic, and hence long time averages can be replaced by averages over the ensemble of all orientational configurations. Such evolution is hence described with the models of statistical mechanics, thus recovering a description familiar from the statistical mechanics of spin Hamiltonians. In short, snapshots of the system are taken with a resolution time $\tau$, with the electrons being described in the LSD approximation in the non-equilibrium state corresponding to the observed orientations of the local moments. The time evolution of the local moments is then described using a classical spin Hamiltonian. A Curie-Weiss law and an estimate of the Curie temperature $T_{Curie}$ = 1250 K is obtained on the basis of the fully itinerant theory [176]. The latter constitutes a substantial advance compared to the Stoner-Wohlfarth model in which the entropy is due entirely to thermal production of electron-hole pairs. In addition, the itinerant theory incorporates the orientational fluctuations (random orientation of sites) as another source of entropy, generally left out in the straightforward generalization of spin-polarized band theory to finite temperatures. It is this orientational entropy that is responsible for obtaining a Curie-Weiss law familiar from the statistical mechanics of spin Hamiltonians.

### 3.3 The magnetism in pnictides: the importance of short time scales characteristic of electron dynamics

The case of metallic Fe is very insightful since it illustrates at best the importance of electron dynamics in understanding magnetism in systems hosting itinerant electrons. First, it shows that electron itinerancy and local spin moments are not mutually exclusive. The example of metallic Fe clarifies that in general the dual itinerant/localized character of electrons does not reflect a simple partition of electrons into localized and/or itinerant ones to which different experiments are sensitive. A clear separation between magnetic and translational degrees of freedom does not occur since both magnetism and electron itinerancy originate from *d*-electrons. The occurrence of different time scales and their distinct separation constitute the rationale for understanding how local moments originate from itinerant *d*-electrons, some of which also happen to participate in the Fermi-surface. Local moments form after averaging out the quantum fluctuations, while on a much longer time scale $\tau_{SW}$ they exhibit a slow motion in which they can change their orientation. This is the Curie-Weiss moment observed in ferromagnetic systems above $T_{Curie}$, with thermal fluctuations described well by spin-Hamiltonians. These considerations allow explaining how spin Hamiltonians can be derived from first principles while fully dealing with itinerant electrons. Put differently, the possibility of fitting spin waves dispersions with spin-Hamiltonian does not automatically imply that electrons cannot be itinerant.

Another important message suggested by the case of Fe metal is that fluctuations of the SM are necessary in order to provide a set of accessible low energy states and additional entropy necessary for the accurate description of the magnetic-to-paramagnetic phase transition. In the case of Fe metal, these fluctuations are the orientational fluctuations of the moments, generally left out in the straightforward generalization of spin-polarized band theory to finite temperatures. As illustrated clearly by Hubbard [173,174], only by taking into account the directional and amplitude fluctuation of the local moments (or the related “exchange field”) the itinerant electrons couple to, can one understand the relatively low transition temperature with a large exchange splitting of the band structure in metallic Fe.

Clearly the physics of Fe-HTSC is different from that of metallic Fe, but some of the considerations outlined above apply to the pnictides as well. Contrary to Fe metal, the pnictides do not exhibit Curie-Weiss behavior. The Curie-Weiss behavior is normally obtained from Heisenberg Hamiltonians as the result of a mean field approximation, which suppresses entirely fluctuations and any correlation effects between moments. The latter can either be coupled directly or interact via their common coupling to other spins. The lack of Curie-Weiss behavior is thus indicative of the occurrence of strong fluctuations, which are neglected in mean field theories.

The presence of strong fluctuations is also revealed by the discrepancy in the magnitude of the SM measured by different experiments. Specifically, we have discussed above how the magnitude of the SM increases corresponding to shorter time scales of the measuring techniques. This discrepancy in the magnitude of the SM between the fast ($\approx 10^{-16}$ s - $10^{-14}$ s) and slow ($10^{-8}$ s - $10^{-6}$ s) measurements is due largely to the occurrence of quantum fluctuations, to which only fast measurements are sensitive. Soft x-ray spectroscopies such as XES and PES are fast probes that allow the measurements of SM fluctuating on time scales as fast as $10^{-16}$ s.

This phenomenology indicates the existence of different energy scales corresponding to how rapidly the system is sampled. These different energy scales correspond to different time limits of the dynamical response of the system: A large ($\approx$ eV) energy scale, indicative of the fast quantum fluctuations, and a small ($\approx$ 1-10 meV) energy scale, which corresponds to dressed interactions forming over a longer time scale. These energy scales manifest in the magnetic response of the system as an instantaneous magnetic moment $m_{inst}$, which correspond to the short time limit of the magnetic susceptibility $\chi''(t = 0)$, the so called dynamical spin susceptibility, and a screened magnetic moment, which corresponds to the static spin susceptibility [136].

Quantum fluctuations manifest directly in fast experiments with a short time constant $\approx \tau_F$, and thus involving large energy transfer. This is the case for example of the PES spectra, which consist is a collection of snapshots of the system taken on the fast time scale of the photoemission process, $\approx 10^{-16}$ s. The values of the local SM extracted from the analysis of the PES Fe 3*s* spectra are thus representative of the system sampled over extremely short time scales characteristic of electron dynamics. Also the lineshape of the Fe 3*s* spectra is indicative of the occurrence of quantum fluctuations, as discussed elsewhere [160]. The analysis of the Fe 3s core level PES spectra thus provides the values of the instantaneous ($10^{-16}$ s) local SM $m_{inst}$. On the contrary, conventional magnetic experiments average over fast quantum fluctuations since they probe the system on time scales much longer than $\tau_F$, with consequent low-energy transfer. The time scale of Mössbauer, NMR and $\mu$-SR measurements are typically $\approx 10^{-8}$ s - $10^{-6}$ s, practically static compared to the time scale of electron dynamics. They measure a screened moment which is strongly reduced as compared to the instantaneous local SM $m_{inst}$ [136].

In essence, Fe-HTSC illustrate the generic feature of magnetism in real materials hosting itinerant electrons, quite distinct from what described in itinerant-only models. In real itinerant magnetic materials, large local moments are always present, but might fluctuate at rather fast time scales.

The presence of local moments and itinerant electrons demonstrated by several experiments emphasizes the deficiencies of some mainstream theoretical approaches in describing the physics of Fe-SC compounds. The significance of local Hund's coupling, revealed by the experimental finding of large fluctuating local moments probed on fast ($\approx 10^{-16}$ s) time scales, highlights the inadequacy of itinerant-only pictures, i.e. those theoretical efforts in describing the magnetism in Fe-HTSC as Fermi surface nesting-driven. The perturbation treatment of the electronic structure in these studies is only reliable with weak interactions, in which case the physics is dominated only by the low-energy Hilbert space in the one-particle channel, that is, near the chemical potential. Therefore, such treatments will not properly capture the rich correlated behavior of the itinerant electrons and their strong interplay with the local moments (higher energy objects in the one-particle Hilbert space).

The occurrence of this interplay, besides being a characteristic feature of magnetic systems hosting itinerant electrons, is suggested by the experimental observation of substantial reduction of the SM upon doping. Specifically, the analysis of the Fe 3s PES spectra revealed large fluctuating SM amounting to 2.1 $\mu_B$ in $SrFe_2As_2$ and 1.3 $\mu_B$ in CeFeAsO that decreases to 1.35 $\mu_B$ and 0.9 $\mu_B$ in the optimally doped samples. A significant reduction of the SM is thus found comparing the 122 parent compound with the 1111 parent compound. Moreover, for both the 122 and the 1111 compounds, the SM decreases substantially in both families on going from the parent to the optimally doped samples (cf. Fig. 13), while changes as a function of temperature are less significant, indicating that the fluctuations associated with the reduction of SM are mostly quantum in nature. This phenomenology is not compatible with a local-only nature of the SM, as the local properties of the Fe ion against doping or materials type cannot change as much to justify the $\approx$ 40% reduction of the SM. On the contrary, these observations reveal the important role played by the itinerant electrons in mediating the magnetism of the pnictides via interaction with the SM. The authors of ref. [160] have discussed how the reduction of the measured SM against doping and material type can be rationalized as a consequence of an increase of the kinetic energy gain, achieved by spreading out the spatial distribution of the fluctuating spins, with the wavefunctions described as spin-polarized Wannier orbitals, onto multiple atomic sites. The fluctuations associated with the large values of the SM reflect the strong competition between the AFM super-exchange interaction among the local SM, and the kinetic energy gain of the itinerant electrons in the presence of a strong Hund's coupling [148,154,160].

As pointed out above, Hubbard and Hasegawa were among the first to propose an amalgam of localized and itinerant models when studying the magnetism in metallic Fe [173,174,175]. They pointed out that the motion of the itinerant electrons and the configurations of the exchange fields, entities essentially proportional to the local SM of atoms, are influenced reciprocally in a self-consistent fashion. In a context specific to the pnictides, it has been discussed how the interaction between the SM is mediated by the itinerant electrons in a self consistent fashion thanks to the provision of additional degrees of freedom such as the low electron kinetic energy and a two-fold orbital freedom, i.e. the degeneracy of the $d_{xz}$ and $d_{yz}$ orbitals [148,154,177,178]. These degrees of freedom are the counterpart of the orientational fluctuations of the magnetic moment in metallic Fe: In the pnictides, the kinetic energy gain and the two-fold orbital degeneracy provide a set of accessible low energy states and additional entropy which add significant flexibility to the system to fluctuate and readjust self-consistently via the interaction of the itinerant electrons with different local magnetic correlations. This interaction provides an effective mechanism for the itinerant electrons to mediate the coupling between the local moments. Electron itinerancy is thus crucial for the magnetism in the pnictides. Consequently, Heisenberg-like local-moment pictures containing a fixed coupling between the local moments result to be inadequate, as they fail in capturing the crucial role of itinerant electrons in mediating the coupling between the SM.

A theoretical description of the pnictides necessitates a more in depth understanding of correlated metals under the influence of strong coupling to local moments. The challenge facing theory is the description of the details of the self-consistent interaction between itinerant electrons and instantaneous local moments in different pnictides. It is not clear at this stage which of the available theoretical approaches can at best describe the physics of the pnictides. Although it has been argued here that the values of Coulomb type electron correlations are moderate at most, the possibility that such correlations exhibit a momentum dependence cannot be dismissed. If this were the case, this occurrence would not be identified in integrated x-ray spectroscopy experiments, which on the contrary provide an estimate of the electron-electron interactions integrated in momentum space and generally without orbital resolution. It has been suggested that intermediate values of the Hubbard U need to be explored in more detail as they might reveal paradigms not previously explored [34]. It is nonetheless expected that results based exclusively on Hubbard models alone would not be able to capture properly the interplay between local moments and itinerant electrons.

A valid alternative could be offered by Spin-Fermion models. The essence of these models is to consider local moments which, besides being coupled to each other directly, are also coupled to itinerant electrons. The basic Hamiltonian would then be written as

$$H = \sum_{i,j} t_{i,j}^{\alpha,\beta} d_{i\alpha}{}^{\dagger} d_{j\beta} + J_I \sum_{i} \hat{\vec{s}}_i \cdot \vec{S}_j + J \sum_{i,j} \vec{S}_i \cdot \vec{S}_j$$

, where $i$ and $j$ denote lattice sites, while $\alpha$ and $\beta$ indicate orbital occupancy. The first and third term describe the kinetic energy of the electrons and the direct Heisenberg-like coupling between local moments, respectively. The coupling between the local moments $\vec{S}_j$ and itinerant electrons with spins $\hat{\vec{s}}_i$ is described in the second term. The models assume that the on-site spins $\vec{S}_j$ are localized, formed by more bound electrons not participating in transport. The interaction of the local moments with the itinerary electrons of spin $\hat{\vec{s}}_i$ can be studied by Monte Carlo methods [179,180].

Another point of view considers the pnictides as Hund's metals [181]. This term was first introduced by Yin et al. [182] to designate materials which, despite showing signatures of correlations, are multi-band metallic systems not in close proximity to a Mott-insulating phase. For these systems, the physical origin of the correlations is not the Coulomb repulsion of electrons in the same orbital, but rather Hund's coupling, the intra-atomic exchange interaction resulting in electrons with parallel spins in different orbitals. Remarkably, correlation effects can occur in itinerant systems even if the bandwidth is significantly larger than the Hund's coupling energy scale. The effect of Hund's manifests both on a high and a low energy scale. On a high energy scale, it increases the effective Coulomb repulsion for a half-filled shell in an isolated atom, while it lowers it for all of the other fillings. On a low energy scale, Hund's coupling lowers considerably the energy scale characteristic of the screening of atomic degrees of freedom. The filling of the atomic shell is found to be an important parameter. For non-half-filled shells, Hund's coupling drives the system away from the Mott transition, but concomitantly it makes the metallic state more correlated by lowering the quasiparticle coherence energy scale [181]. Accordingly, the presence of strong correlations no longer implies necessarily proximity to a Mott phase: in multiband itinerant systems, Hund's coupling can induce strong correlation effects even for modest values of the on-site Coulomb repulsion and for large bandwidths. Modest value of the Coulomb repulsion and strong intra-site exchange correlations are not mutually exclusive, given that while the value of the Coulomb repulsion can be lowered significantly in a solid with respect to the isolated atom, for the exchange interaction the reduction amounts to only 20-30% [181,183]. For an in-depth explanation of the correlation effects due to Hund's coupling in itinerant system the reader is referred to the review by Georges, de Medici and Mravlje [181]. This reference also provides a general introduction to the method of Dynamical Mean Field Theory (DMFT) which, as argued by the authors, provides the most appropriate theoretical framework to describe Hund's correlations, as it describes band-like and atomic-like aspects on an equal footing [181]. In contrast to more traditional approaches which describe a solid starting from an inhomogeneous electron gas to which interactions are then added at a second stage, DMFT focuses on the fact that a solid is composed by atoms whose multiplet structure presents a many-body problem, i.e. many-body correlations on each atomic site, which has to be addressed from the start. The electron transfer among atoms in the solid is then addressed by focusing on a single atomic site and describing the rest of the solid as an effective medium that exchanges electrons with the single atomic site [181].

In the context of the pnictides, the reduction of the Drude weight at low energies and its recovery at higher energies (above 8000 $cm^{-1}$) observed from optical measurements in $BaFe_2As_2$ have been interpreted as signatures of Hund's coupling [97]. Haule and Kotliar have emphasized the importance of both correlations and metallic behavior [60,116]. The same authors have proposed a scenario featuring a crossover from coherent itinerant magnetism to incoherent local moment magnetism occurring at a temperature T* [116], but the predicted temperature evolution of the magnetic susceptibility, from Pauli-like to Curie-Weiss-like, has been found not to be in agreement with experiments [31,137]. More recent theoretical work within the DMFT framework is particularly noteworthy. The overestimation of the size of the moment by LDA suggests the possibility that dynamical effects are at work. Focusing on the high energy scale (short-time scale), DMFT+LDA calculations by Hansmann et al. have indicated that that the value of the moment is rather large (S ~ 2), but the screening of the metallic environment causes it to decay very rapidly, within a few fs [136].

Neutron scattering measurements of the dynamical spin susceptibility in Ni-doped $BaFe_2As_2$ has revealed smaller moments (S = 1/2) [159]. The dynamical spin susceptibility appears to be affected by doping only in the low energy sector, i.e. for energies less than ~ 80 meV (cf. Fig. 14a). Remarkably, LDA+DMFT calculations of the spin susceptibility in absolute units have reproduced the results. A comparison of these results with Random Phase Approximation (RPA) calculations indicated that the latter places the peak of the spin susceptibility at energies approximately one order of magnitude larger than the 200 meV value found by LDA+DMFT, which corresponds to fluctuation on time scales of the order of 20 fs [159].

In the author's view, these results emphasize further the importance of electron dynamics effects in the physics of the pnictides. Given the similarity of the Co- and Ni-doped $BaFe_2As_2$ systems, the neutron results reported in ref. [159] can be compared to the PES results reported in ref. [160]. The neutron measurements show small SM corresponding to S = ½, with the high energy sector of the dynamical magnetic susceptibility not being affected by doping. On the contrary, the PES measurements indicate a much larger moment (S = 2) in the $BaFe_2As_2$ parent compound which decreases by 40 % in the optimally doped compound. It is possible that the small SM measured by neutron scattering is a consequence of the impossibility of disentangling completely the scattering signal from the background while performing the integration of $S(\vec{Q},\omega)$, resulting in loss of spectral weight. Equally important is the fact that INS and PES sample the system on different time scales: An integration of the INS data up to ~ 300 meV correspond to time scales ~ 10 - 15 fs, at least one or two orders of magnitude slower than those in PES experiments. Since PES samples the system on time scales shorter than $10^{-15}$ - $10^{-16}$ s, the PES data reveal the existence of large SM on extremely short time scales that cannot be probed by INS. In fact, the maximum of the energy range (~ 300 meV) probed by INS experiments imposes a limitation to sampling the system for time scales below ~ $10^{-14}$ s. The emerging picture is consistent with the scenario proposed by Hansmann et al. [136], that is, large moments form on extremely short time scales (a few fs), but they decay very rapidly due to screening in the metallic environment. The INS data are representative of the systems after screening processes begin to be effective, providing a value of the SM much lower than that of the instantaneous value $m_{inst}$ measured by PES. Rigorously speaking, referring to the existence of instantaneous SM, i.e. SM on short time scales which coincide with the time scale $\tau_Q$ typical of quantum fluctuations, is not correct according to the description of moment formation reported in the literature of metallic Fe [172]: Spin moments form after averaging out the quantum fluctuations, and thus SM form on time scale larger than $\tau_Q$. Nevertheless, the instantaneous value of the SM provided by the PES experiments is meaningful, as it provides a quantitative estimate of the instantaneous local magnetic moments (Hund's coupling) that the itinerant electrons interact with, allowing the system to readjust self-consistently. This process happens fast, within ~ 10 fs, as suggested by the fact that INS data show signatures of the screening of the instantaneous SM for energy of 200-300 meV (i.e. 10-20 fs, cf. Fig. 14b). The marked reduction of the instantaneous SM upon doping emphasizes the sensitivity of the self-consistent readjustment of the system to different carrier concentrations on fast time scales (< 10 fs).

A theoretical description of the pnictides necessitates a more in-depth understanding of the details of the self-consistent interaction between itinerant electrons and instantaneous local moments for different doping levels, with particular attention to be paid to the extremely short time scale ~ 1 fs, that is, energy scales ~ eV as probed in x-ray spectroscopy experiments. As shown in Fig. 14b, the eV energy scale is where the RPA calculation places the maximum of the spin susceptibility. The RPA approximation contemplates essentially only particle-hole pair excitations, which are higher in energy than the set of accessible lower energy states provided by other degrees of freedom such as, for example, the kinetic energy gain and the two-fold orbital degeneracy as proposed in refs. 148,154,177,178. Experimental confirmation that these are the very degrees of freedom responsible for the provision of lower energy states is necessary to elucidate the mechanisms according to which itinerant electrons interact with the instantaneous spin moments and mediate the coupling among the latter. Anyhow, irrespective of what the mechanisms may be, the occurrence of different energy scales corresponding to different limits of the dynamical response of the system leads credence to the fact that low energy states and additional entropy add significant flexibility to the system to fluctuate and readjust self-consistently via the interaction of the itinerant electrons with the spin moments. This is analogous to the case of metallic Fe, in which the fluctuations of the moments in directions constitute a set of excited states with much lower energy than the Stoner excitations contemplated in standard spin-polarized band theory. An extension to the eV scale of the calculation results shown in Fig. 14b is promising for reaching a more comprehensive understanding of the time evolution of the renormalization of the instantaneous spin moment $m_{inst}$.

## 4. CONCLUDING REMARKS

In summary, the nature of the electronic correlations and the closely related microscopic origin of magnetism in Fe-based pnictides high temperature superconductors (HTSC) have been reviewed. The strength of electron correlations and the microscopic origin of magnetism in Fe-based pnictides HTSC have been the cause of an intense debate since the start of the field in 2008. According to many experimental studies, the pnictides exhibit an itinerant electron character which is incompatible with the occurrence of strong local Mott-Hubbard type correlations that for example characterize the cuprates. On the other hand, neutron scattering experiments seemed to be well described by a local picture, with local magnetic moments on the Fe sites, although with a much reduced magnitude compared to the theoretical predictions. The occurrence of localized and itinerant characters of electron spins suggested by different experiments has been at the beginning often interpreted and presented as an axiomatic dilemma between two extremely different theoretical viewpoints: The first one described the electrons as localized, while the other viewpoint identified the Fe-HTSC as itinerant magnets.

Several experimental and theoretical works proposing that the Fe *d* electrons have both localized and itinerant characters appeared as early as 2009. Neutron scattering studies have suggested the importance of both localized and itinerant Fe *d* electrons, with similar conclusions been proposed on the basis of optical spectroscopy experiments. On the theoretical front, several works have indicated the necessity to attain descriptions beyond a fully localized or completely itinerant perspective. Nonetheless, pnictide materials were observed to be uniformly metallic throughout their doping/pressure phase diagrams, with a high density of states at the Fermi level, without signatures of strong local Mott-Hubbard type electron correlations, and with a nested Fermi surface compatible with an enhancement of the particle-hole susceptibility and spin-density-wave order. Conventional density functional methods, which typically fail for correlated systems, were shown to capture many of the electronic properties. Taken together with the small values of the magnetic moments both in the ordered and paramagnetic phases, these facts have shaped in the community the belief that, in the pnictides, the anti-ferromagnetic order does not stem from exchange interactions between local magnetic moments, but instead has its origin in the Fermi surface nesting, with Coulomb correlations not playing a significant role in the formation of the magnetic ordered states.

New results provided in the last couple of years based on inelastic neutron scattering, x-ray emission, and photoemission experiments indicate that the pnictides host a more complex physics than originally anticipated. These experiments have unveiled the presence of large spin moments fluctuating on fast, i.e. sub-picosecond, time scales in magnetically ordered, paramagnetic, and superconducting phases. These observations are important for several reasons. First, the presence of local moments in the paramagnetic phase is incompatible with theories relying exclusively on Fermi surface instabilities. Second, the large values of the Fe spin moment indicate the occurrence of rather strong on-site exchange correlations (Hund coupling) that fosters the electrons in the Fe *3d/4s* shells to align parallel to each other. Hund's coupling can provide a rationale for the occurrence of large local moments and renormalizations and shifts in spectral weight within the Fe *d*-bands while maintaining an itinerant electron character, as advocated by initial theoretical work based on dynamical mean field theory. Third, a comparison of the magnitude of the spin moment as measured with inelastic neutron scattering, x-ray spectroscopy, and more conventional magnetic techniques reveals that the values of the spin moment increase when measurements are performed on faster time scales. This phenomenology indicates the existence of different energy scales corresponding to different time limits of the dynamical response of the system: A large (≈ eV) energy scale, indicative the fast quantum fluctuations and manifesting as the short time limit of the dynamical spin susceptibility $\chi(t = 0)$, i.e. the instantaneous spin moment, and a small (≈ 1-10 meV) energy scale, which corresponds to the static spin susceptibility, i.e. the renormalized spin moment. The occurrence of different time scales and their distinct separation constitute the rationale for understanding how local moments originate from itinerant *d*-electrons.

Electron itinerancy is another important aspect of the physics at play in the pnictides. Hubbard bands have not been found, and density functional theory calculations appear to be in good agreement on the ~ 1-10 eV energy scale probed by x-ray spectroscopy experiments, which have not shown spectral signatures commonly found in strongly correlated systems. These experiments indicate that Hubbard-type correlations are weak or moderate at most, but certainly not negligible. In fact, a renormalization of ~ 2 is found to be necessary in order to obtain an agreement between the measured and calculated band structure within 1-2 eV from the Fermi level. Photoemission experiments have provided a direct estimate of the hole-hole Coulomb repulsion $\mathcal{U}$ via measurements of the core-valence-valence Auger transitions. In units of the electron bandwidth $W$, these measurements provide values of $\mathcal{U}/W$ in the 0.3 - 0.4 eV range, indicating that the interaction between holes in the valence bands of the studied pnictide compounds is highly screened. As discussed, values of $\mathcal{U}$ cannot be directly identified with the value of the parameter $U$ in model calculations of the Hubbard type, but rather provide an upper bound to the values of $U$ in theoretical descriptions of phase diagrams that vary with the quantity $U$. The role of itinerant electrons appears to be very important for understanding the magnetism in the pnictides. Specifically, it has been proposed that the provision of additional degrees of freedom such as the low electron kinetic energy and the $d_{xz}$ - $d_{yz}$ orbital degeneracy provide a set of accessible low energy states and additional entropy which add significant flexibility to the system to fluctuate and readjust self-consistently via the interaction of the itinerant electrons with the local spin moments, providing an effective mechanism for the itinerant electrons to mediate the coupling between the local moments. Experimental confirmation of these proposals is necessary to elucidate the mechanisms according to which itinerant electrons interact with the instantaneous spin moments and mediate the coupling among the latter.

Other fundamental questions remain to be elucidated in order to attain a sound understanding of the physics shaping the macroscopic properties of the pnictides, including in particular high temperature superconductivity and the its possible relation to the magnetism and magnetic order. The occurrence of electron itinerancy, the presence of strong exchange correlations, revealed by the detection of large local spin moments, and the absence of strong Hubbard type correlations lead credence to the fact that the pnictides can be consider Hund's metals, a term coined to designate materials which, despite showing signatures of correlations, are multi-band metallic systems not in close proximity to a Mott-insulating phase. Hund's coupling can induce strong correlation effects in multiband itinerant systems large bandwidths even for modest values of the on-site Coulomb repulsion. The Hund metals framework provides a rationale for the presence of both strong intra-site exchange correlations and modest Coulomb repulsion due to the different degrees of screening of these interactions in a solid environment. Studies based on dynamical mean field theory have shown that the latter has been very successful in capturing the physics of some of Hund metal systems, including some aspects of the pnictides. As discussed, the values of the hole-hole repulsion $\mathcal{U}$ as provided by photoemission experiment suggest that Hubbard type correlations are strongly screened. Nevertheless, it is important to point out that values of $\mathcal{U}$ are provided by experiments without momentum space and, generally, orbital resolution. On the contrary, the values of the parameter $U$ in calculations of the Hubbard type strictly depend on specific models and possibly assume different values in specific orbitals, this possibility being not detectable, in general, by the experiments. Hund metals frameworks contemplate the possibility that the strength of the interactions can be orbital dependent, but in general the possible relevance of Hubbard type correlations in the intermediate regime cannot be a priori dismissed and need to be explored. The pnictides present the challenge of understanding in detail the interplay between itinerant electrons and local moments. More specifically, an accurate theoretical description of the pnictides necessitates a more in-depth understanding of the details of the self-consistent interaction between itinerant electrons and instantaneous local moments for different doping levels, with particular attention to be paid to different time scales. Finally, this review has not discussed the important issue of magnetic ordering in Fe-HTSC. A correct description of both the mechanisms leading to the ordering and the ordering direction (the wavevector $\vec{Q}$) for the whole field of Fe-HTSC is lacking. In fact, as mentioned in the Introduction, it is not clear at the moment what is the correct picture embracing the whole field of Fe-HTSC. As mentioned above, in real materials any proper description of magnetism falls on a spectrum having as limits a fully localized picture on one side, and a fully itinerant picture on the other side. Different families such as pnictides and alkali metal iron selenides fall on different parts of the spectrum, as suggested by the magnitude of the Hubbard $U$. Those materials that tend to fall on the localized side of the spectrum, such as the alkali metal iron selenides and chalcogenides, a local moment picture for the magnetism appears to be appropriate. This is consistent with the presence of stronger electron correlations and the values of the SM significantly larger than those in the pnictides. In this case the discussion of magnetic ordering can likely take place by considering the inter-site coupling of local SM. For the pnictides the situation is different; as argued in this Review, the physics of these materials is dictated by the interplay between itinerant electrons and large SM fluctuating on rapid timescales. Any theory hoping to address magnetic ordering in an unbiased way will have to treat the presence of large fluctuating SM interacting with itinerant carriers and the ordering tendencies of the renormalized SM on an equal footing.

In summary, the physics at the heart of the macroscopic properties of pnictides Fe-based high temperature superconductors appear to be far more complex and interesting than anticipated. With the provision of several compounds belonging to different families, pnictides materials offer the opportunity of studying important paradigms for the physics of complex electron systems such as Hubbard type correlations in the intermediate regime, exchange correlations in multi bands metallic systems (Hund metals), and the relation between high temperature superconductivity and magnetism.

## ACKOWLEDGMENTS

The author is grateful to E. Dagotto, P. Dai, T. Egami, A. Eguiluz, M. D. Johannes, S. Johnston, W. Ku, T. Maier, D. Mandrus, M. A. McGuire, A. Moreo, P. Phillips, B. C. Sales, D. J. Singh, M. Stocks, J. Tranquada, P. Vilmercati, H. Weitering, J. Zaanen, and I. A. Zaliznyak for insightful discussions. Special thanks to S. Johnston for a critical reading of the manuscript.

This work was supported by the National Science Foundation, Division of Material Research, grant DMR-1151687.

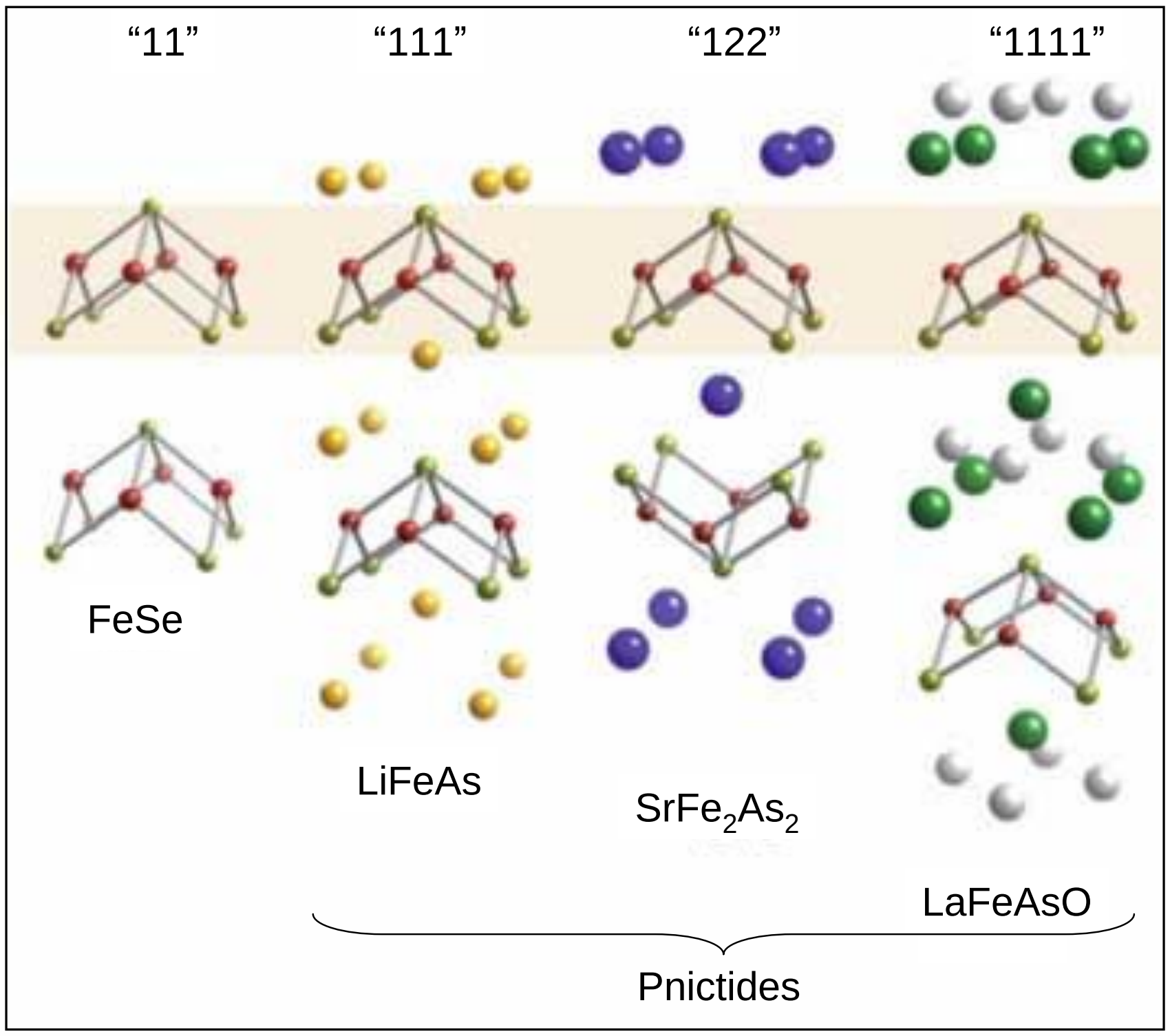

Fig. 1  Crystal structure of the FeSE 11 family and the 111, 122 and 1111 pnictides family. The shaded region indicates the basic structural building blocks, i.e. square planar nets of Fe atoms arranged in a tetrahedral environment. The bond angles and the positions of the Pn/Ch atom above Fe in the tetrahedra display a correlation with $T_C$. Adapted from ref. [33].

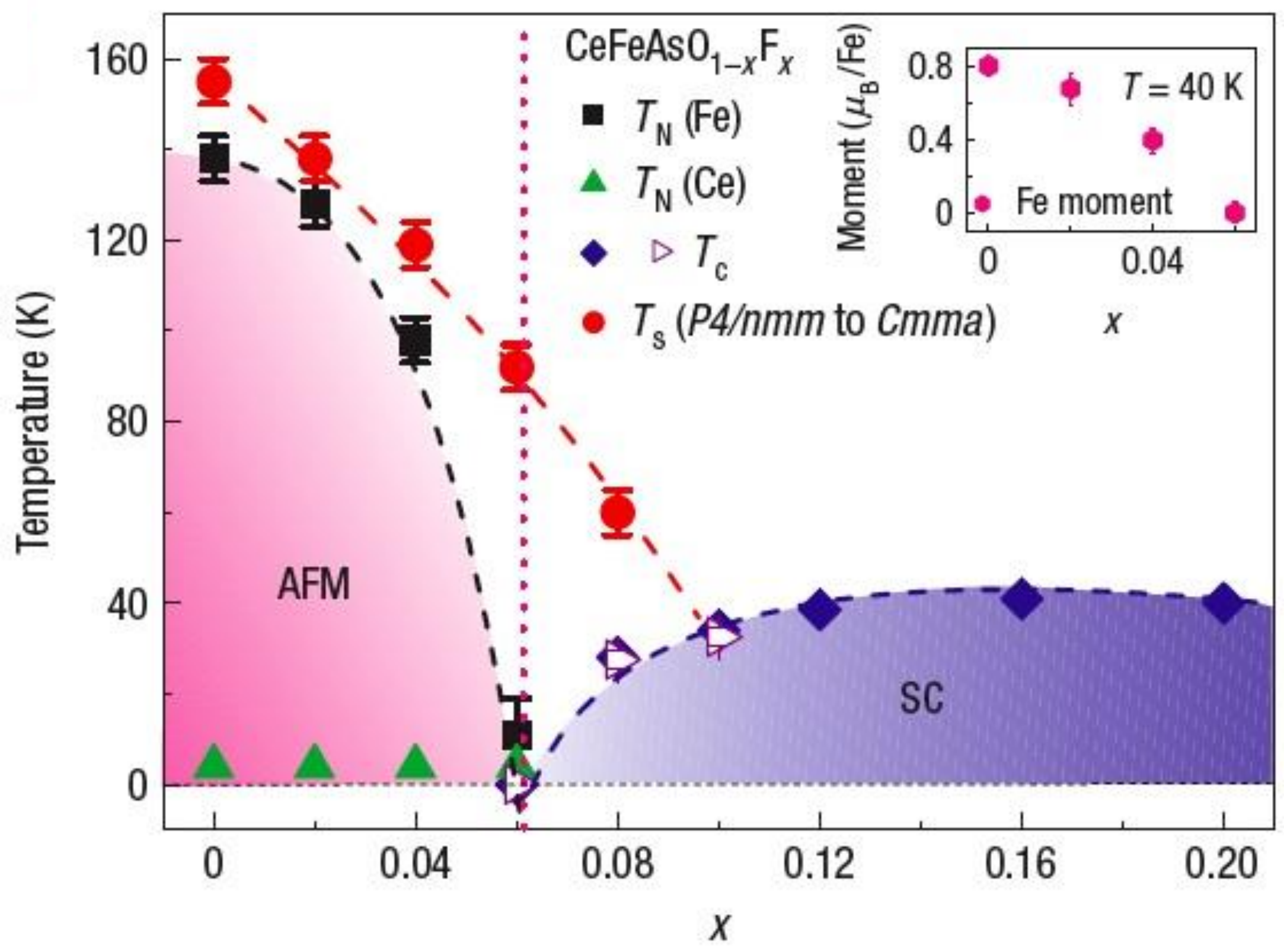

Fig. 2 Phase diagram for the 1111 compound $CeFeAsO_{1-x}F_x$. The presence of the antiferromagnetic region in proximity of the parent compound and the emergence of the superconducting dome for higher doping level is a common characteristic of pnictides compounds. From ref. [128].

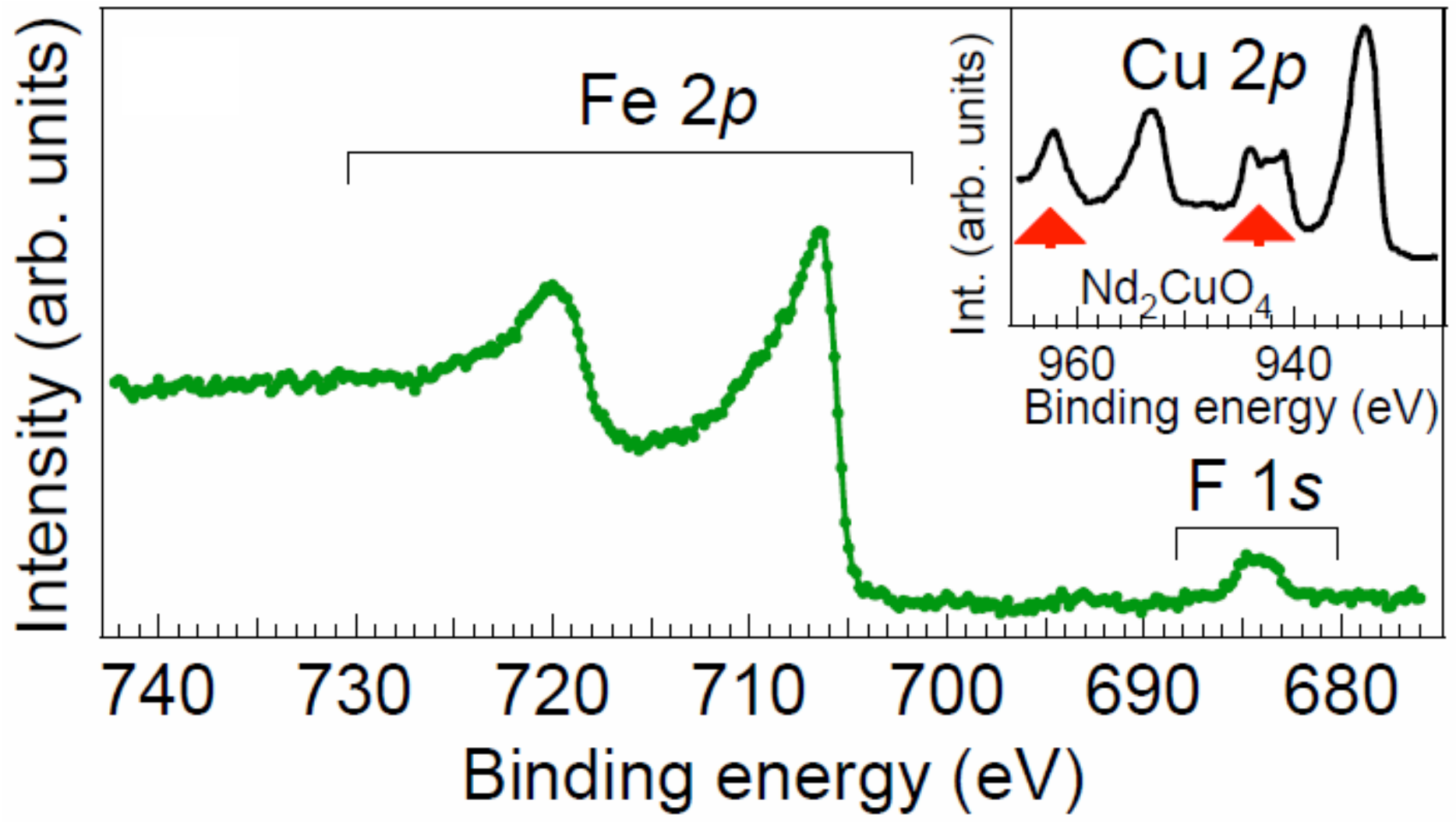

Fig. 3 Fe 2p core level PES spectrum in $CeFeAsO_{0.89}F_{0.11}$. Also visible is the F 1*s* spectrum. The two peaks in the Fe 2p spectrum correspond to the split degeneracy of the 2p manifold into the $2p_{1/2}$ and $2p_{3/2}$ levels due to the spin-orbit interaction. The inset shows the Cu 2p spectrum in the cuprate $Nd_2CuO_4$ for comparison. Satellite structures, indicated by arrows, are present in the Cu 2*p* PES spectrum, but absent in the Fe 2p spectrum. The absence of satellite structures in the Fe PES spectra indicate that the core-hole excitation is completely screened by the Fe states at $E_F$. From ref. [98].

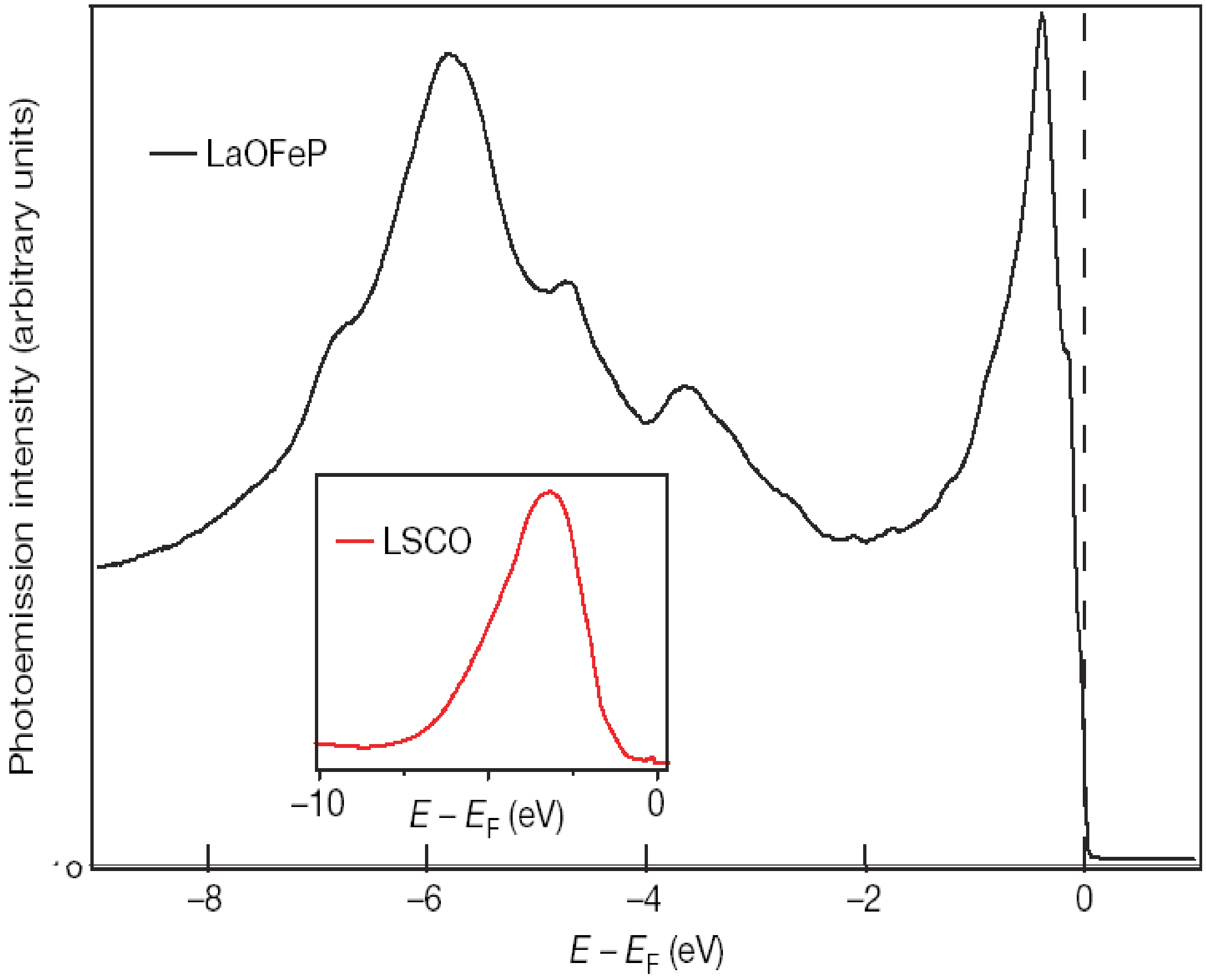

Fig. 4 Valence band measured with PES of the phosphate compound LaOFeP. Note the high intensity of the signal in proximity of the Fermi level ($E_F$), i.e. $E - E_F = 0$, indicating a high density of states. This contrasts markedly the low signal at $E_F$ typical of cuprates, as shown in the inset for comparison. From ref. [91].

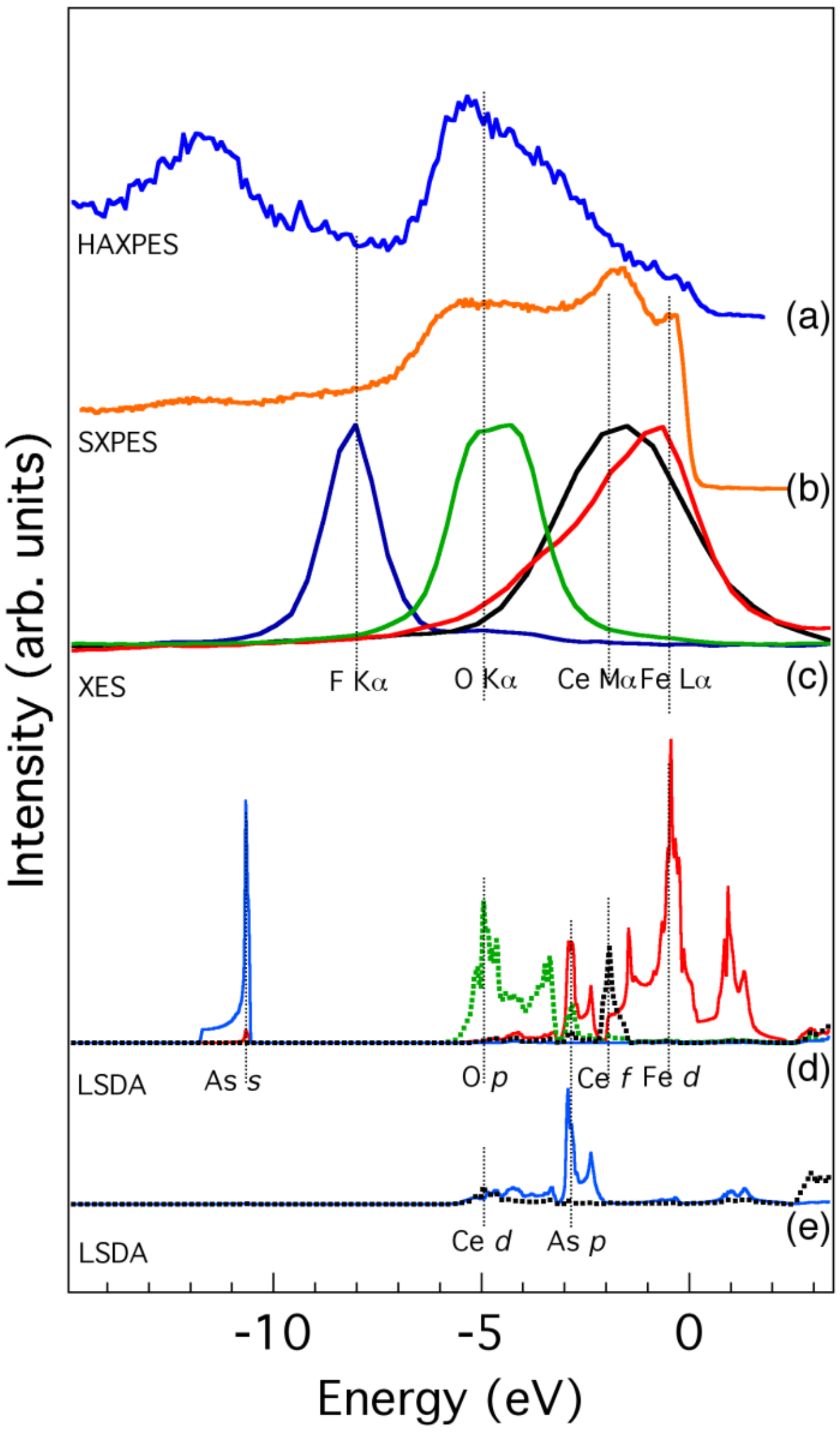

Fig. 5 (a) Valence band high energy (HAXPES) and (b) soft x-ray (SXPES) photoemission spectra measured with photon energy of 7596 eV and 175 eV, respectively, which provide a representation of the occupied total density of states weighted by the orbital cross section. (c) F $K_\alpha$, O $K_\alpha$, Ce $M_\alpha$ shallow core levels and Fe $L_\alpha$ XES spectra measured at room temperature. These near-threshold XES spectra are aligned to a common energy scale with respect to the core binding energies, reference to the Fermi level, enabling the decomposition of the valence band in the (F 2*p*, O 2*p*, Ce 4*f* and Fe 3*d* ) partial density of states components. (d) and (e) Partial As *s*, As *p*, O *p*, Ce *f*, Ce *d* Fe *d* DOS (average of majority and minority states) calculated for a virtual crystal with 10% doping. Note the agreement of the peaks position as provided by experiments and DFT calculations. From ref. [100].

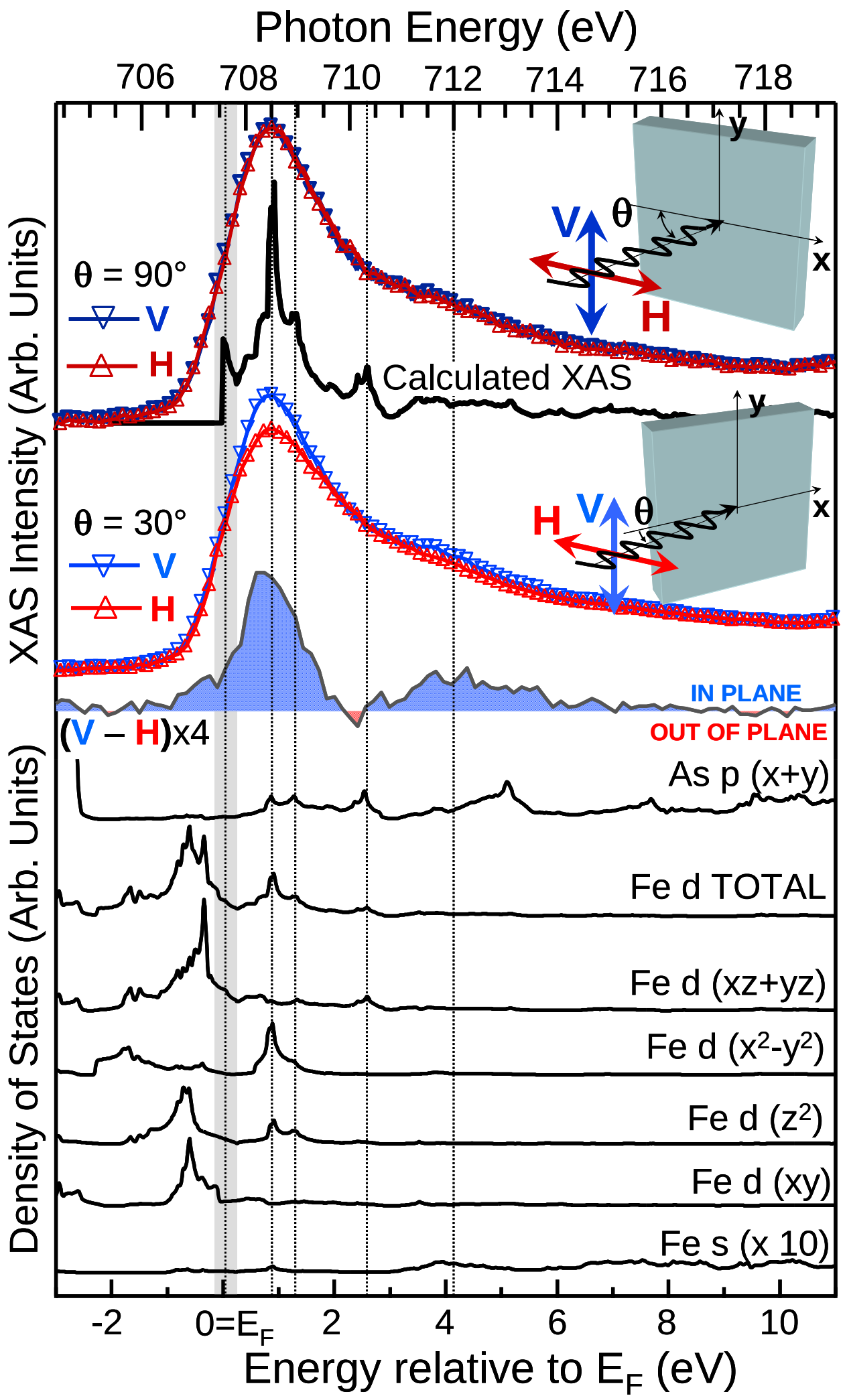

Fig. 6 Fe $L_3$ edge XAS spectra measured in $BaFe_2As_2$ and comparison with theory. The insets show the schematic layout of the experimental geometries. The incident beam is at normal (θ = 90°) and grazing incidence (θ = 25°) to the surface with horizontal (H) and vertical (V) polarizations, as denoted by the double headed arrows. The x and y axes denote the direction of the Fe-Fe bonds. The thick black line denotes the XAS spectrum calculated using orientation averaged matrix elements and DOS calculated with DFT without magnetism. The curve denoted as "V-H" is the dichroic signal obtained by subtracting the H polarization spectra from the V polarization spectra at grazing incidence. The bottom part of the figure shows the orbital projections of the Fe-s and Fe-d p-DOS calculated with DFT. The As $p_{x+y}$ states are also indicated to illustrate the hybridization with the Fe s and Fe $d_{xz+yz}$ states. From ref. [99].

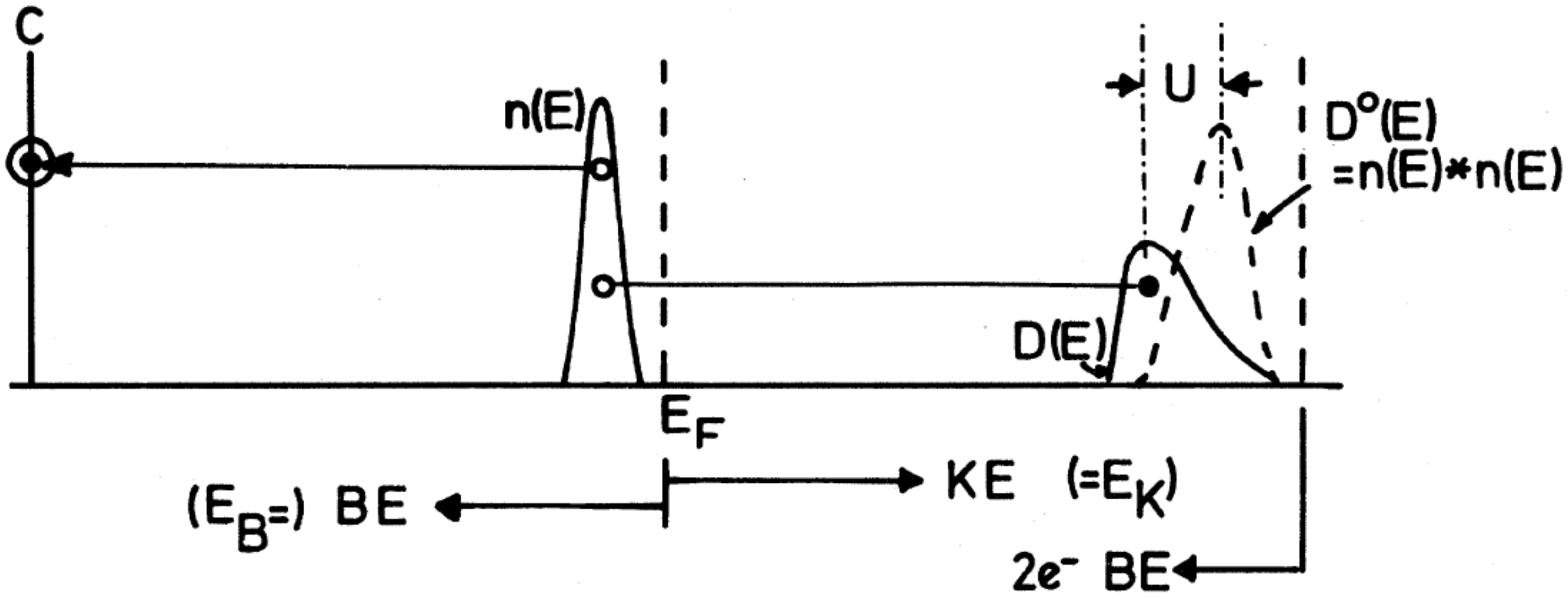

Fig. 7 Schematic diagram for the CVV Auger process. A hole in the core level C is filled by one electron from the valence band, while another electron from the valence band is promoted in the continuum (i.e. the Auger electron). In absence of hole-hole correlation in the two-hole final state, the Auger spectrum resembles the self convolution $D^0(E)$ (dashed line) of the partial density of states n(E). When correlation effects are not negligible, the spectral weight of Auger spectrum shifts to lower kinetic energy (continuous line). The different in the centroids of the measured Auger spectrum and the self convolution $D^0(E)$ provides the value of the screened hole-hole repulsion $\mathcal{U}$. From ref. [107].

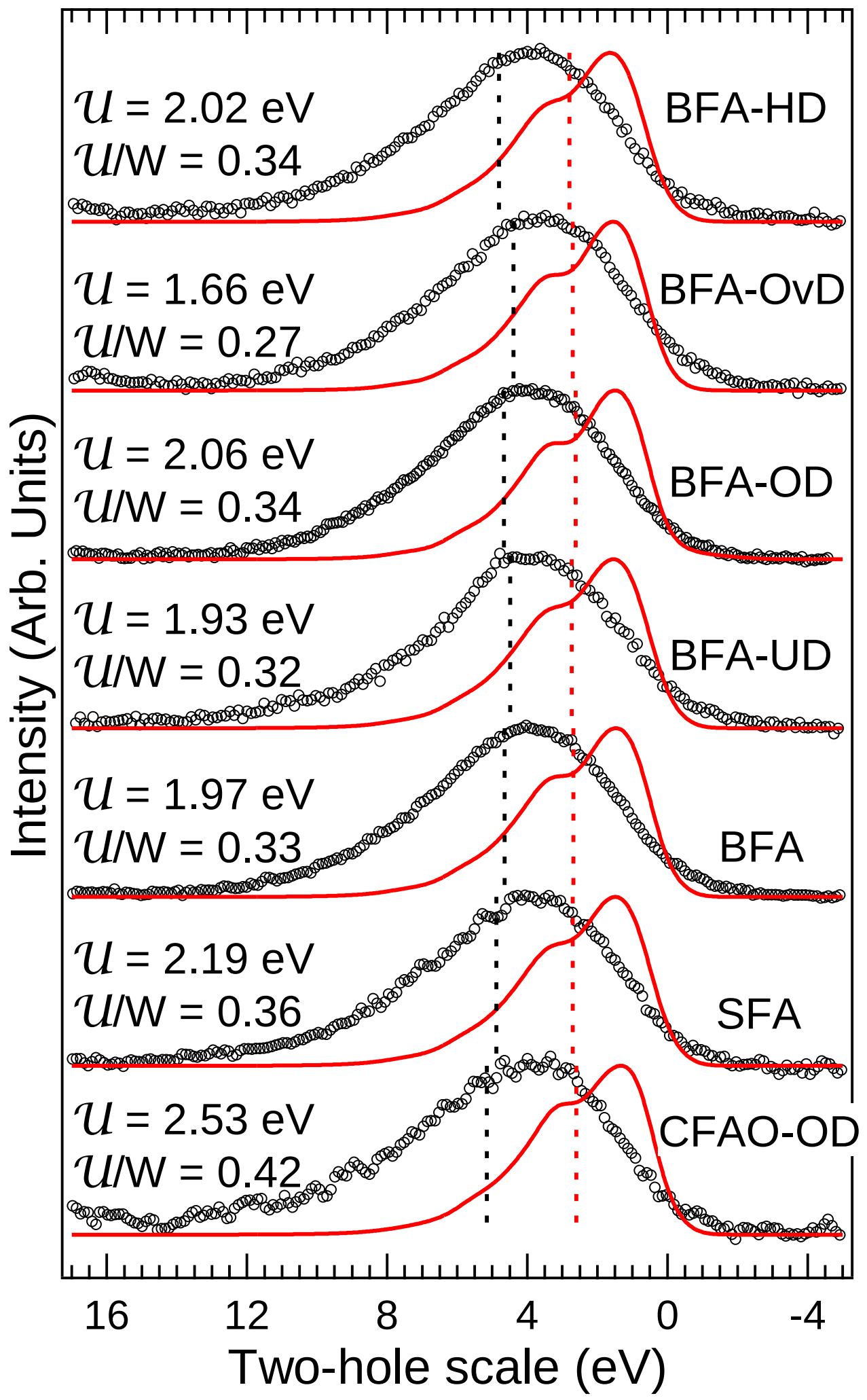

Fig. 8 Determination of the effective hole-hole repulsion energy $\mathcal{U}$ with photon-excited Auger spectra. Fe2$p_{3/2}$VV Auger spectra and calculated $D^0(E)$ are plotted on the two-hole scale for different compounds. The two-hole scale is obtained by subtracting the binding energy of the initial core hole (Fe 2p) from the measured kinetic energy of the Auger spectrum. The dotted lines denote the centroids (i.e. weighted averages) of the Auger spectra and the $D^0(E)$ lineshape. The energy difference between the centroids provides an experimental assessment of the value of the effective $\mathcal{U}$. The measurements were performed on polycrystalline $CeFeAsO_{0.89}F_{0.11}$ (optimally doped, CFAO-OD), and $Ba(Fe_{1-x}Co_x)_2As_2$ and $SrFe_2As_2$ (SFA) single crystals. For the $Ba(Fe_{1-x}Co_x)_2As_2$ system (BFA) the doping levels are $x$ = 0, 6%, 8%, 12%, 22%, corresponding to parent compound (BFA), under-doped (UD), optimally doped (OD), over-doped (OvD) and heavily over-doped (HD), respectively. From ref. [105].

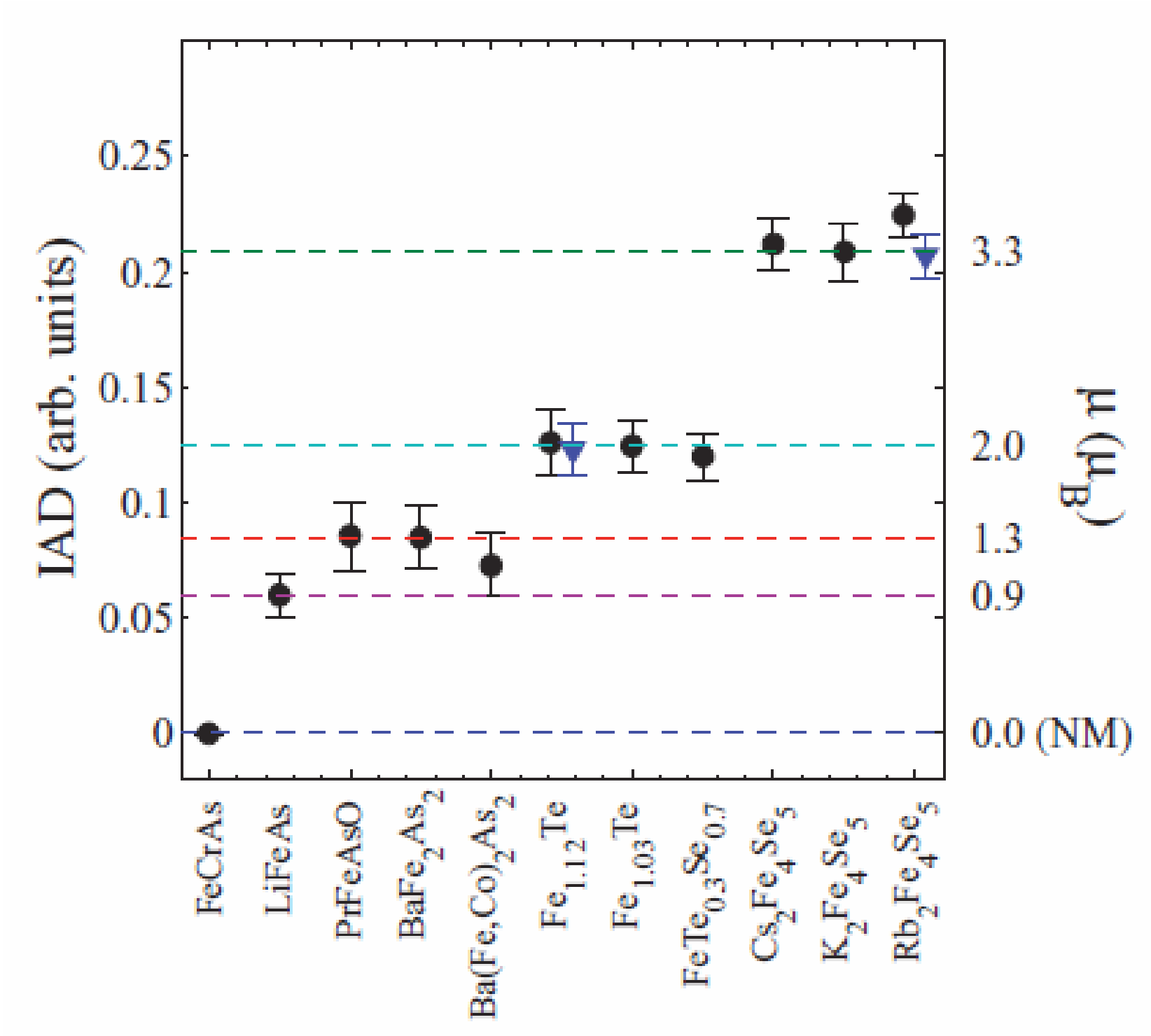

Fig. 9 Values of the magnetic spin moments in the paramagnetic phase (circles) extracted from the XES measurements. The scale on the left side of the plot denotes the integrated absolute difference (IAD) between the measured samples and a reference sample with the same local coordination around Fe, but with Fe ion in the nonmagnetic (S = 0) state, which is necessary for quantitative determination the total local moment from the analysis of the Kβ line. From ref. [126].

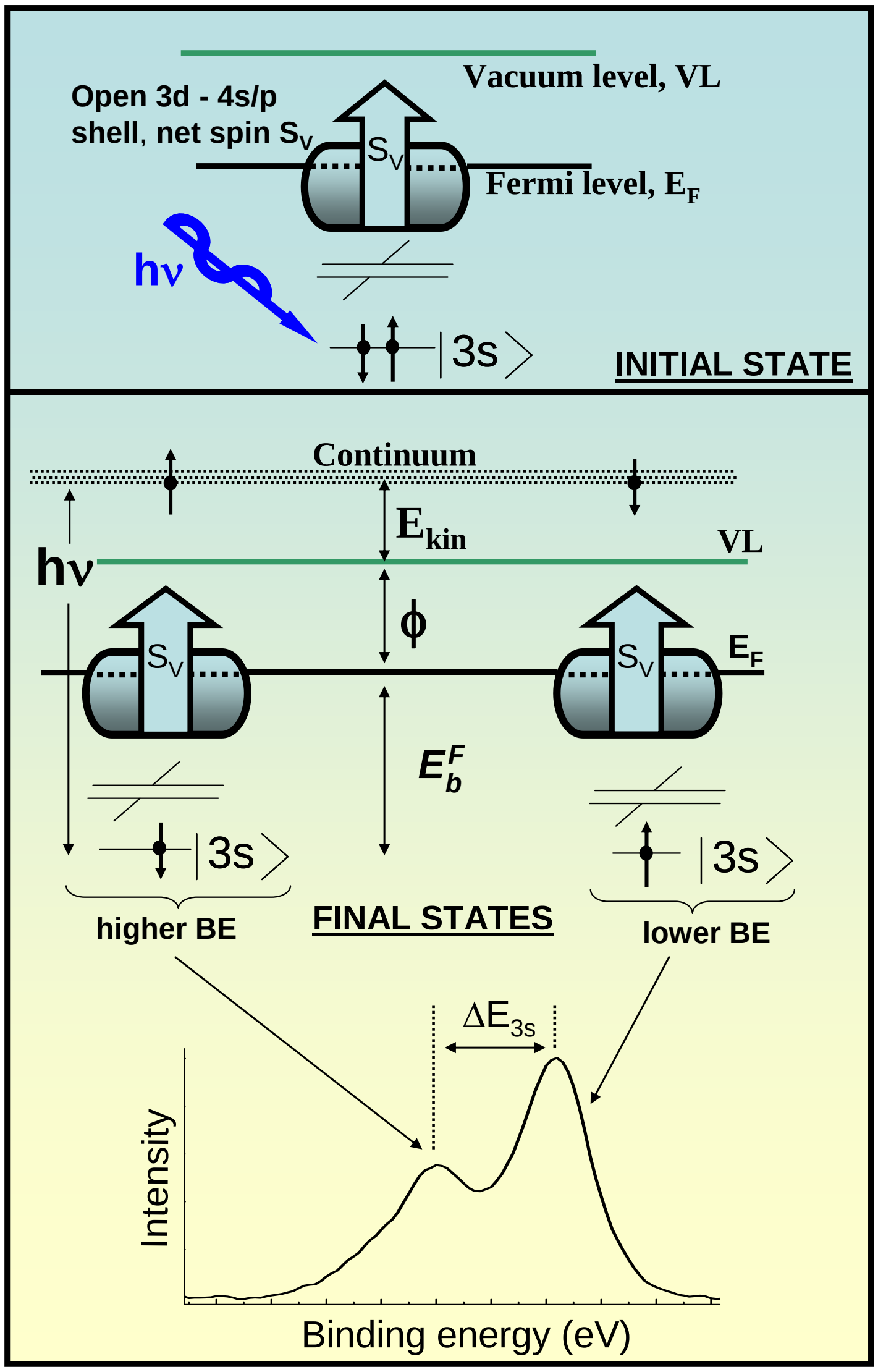

Fig. 10 Schematic layout of the multiplet splitting of the Binding Energy (BE) in 3s core level spectra of transition metals (TM). The upper panel shows the schematic energy levels of a TM atom with an unfilled shell with total net spin $S_V$ formed by electrons in the *TM 3d* and *4s/p* levels. The TM *3s* core levels host two electrons with opposite spins. Upon absorption of a photon of energy $h\nu$, electrons in the *3s* core levels are excited in the continuum above the vacuum level. For a system of *N* particles with ground state energy equal to $E_0(N)$, energy conservation in the photoemission process requires that $h\nu + E^0(N) = E_{KIN} + \phi + E^*(N\text{-}1)$, where $E_{KIN}$, $\phi$ and $E^*(N\text{-}1)$ denote the kinetic energy of the photoelectron, the work function, and the energy of the *N-1*-particle system in the presence of the core hole left behind upon photoelectron emission.

Importantly, the asterisk (*) indicates that the final state of the photoemission process involves in general excited states of the remaining *N-1*-particle system. By detecting photoelectrons of kinetic energy $E_{KIN}$, the photoemission process allows the measurement of the BE defined as $E_B^F = h\nu - E_{KIN} - \phi = E^*(N\text{-}1) - E^0(N)$, where the *F* superscript indicates that the BE is referred to the Fermi level. This expression makes clear that final states *E*(N-1)* of lower (larger) energies are detected at lower (larger) BE. Upon emitting an electron from the *3s* core level, two final states are possible, corresponding to the configurations in which the remaining core *3s* electron is either parallel or anti-parallel to the net spin $S_V$ in the unfilled *3d-4s/p* shell of the TM atom. The exchange energy of the state with parallel spins is lower than that with anti-parallel spins. The multiplet separation $\Delta E_{3s}$ corresponds to the energy difference between these two final states.

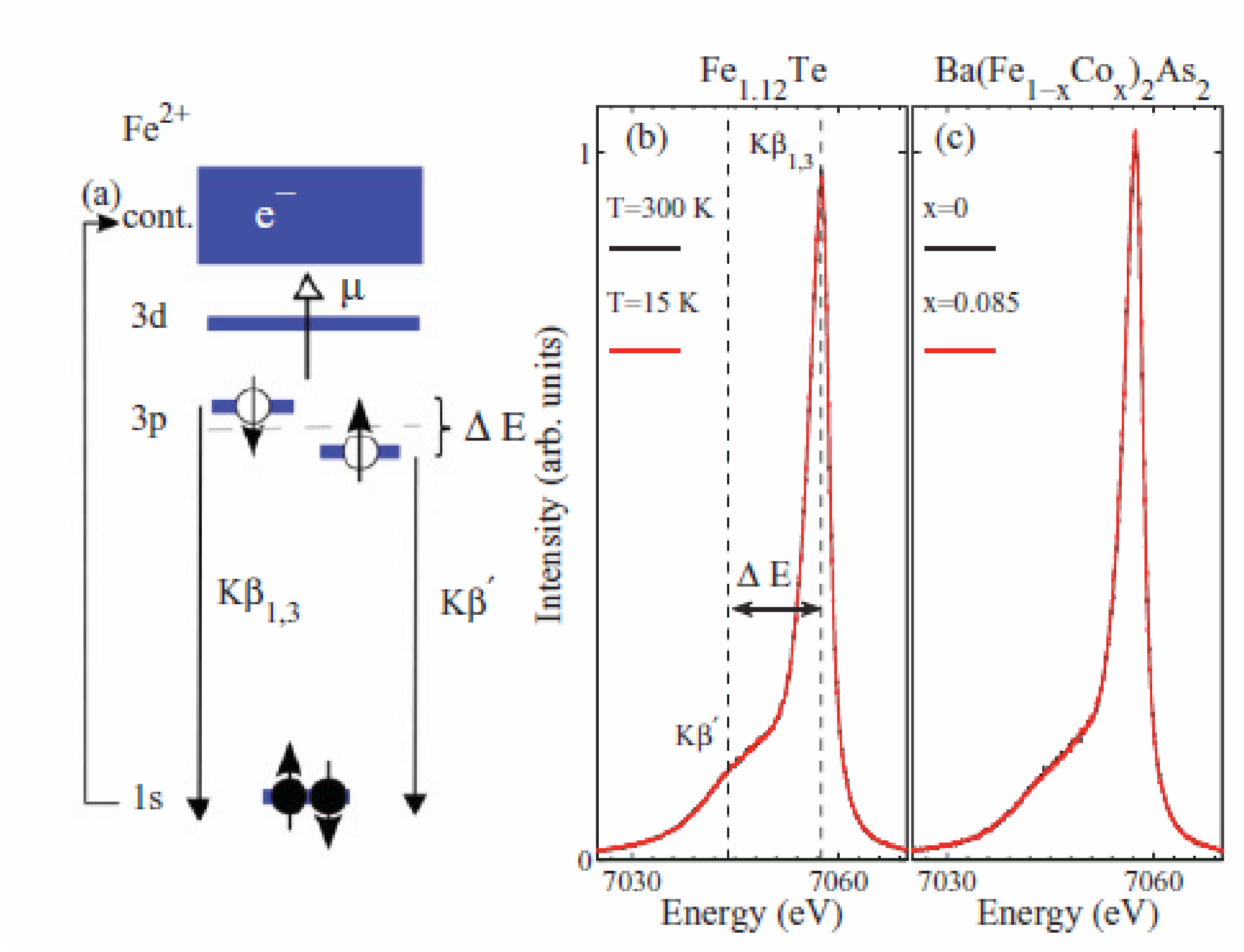

Fig. 11 Schematic diagram of the multiplet splitting mechanism for detecting magnetic moments with XES. (a) Fe K*β* emission process in the atomic limit for $Fe^{2+}$. The spin of the 3*p*

core hole in the final state interacts with the net magnetic moment $\mu$ in the 3*d* valence shell. The energy ΔE denotes the difference in energy of the two final states, i.e. with spins parallel and anti-parallel to the spin moment μ. (b-c) XES spectra of the K$\beta$ emission line for $Fe_{1.12}Te$ and $BaFe_2As_2$. The difference between the two emission lines $K\beta_{1,3}$ and $K\beta'$ correspond to the energy ΔE. From ref. [126].

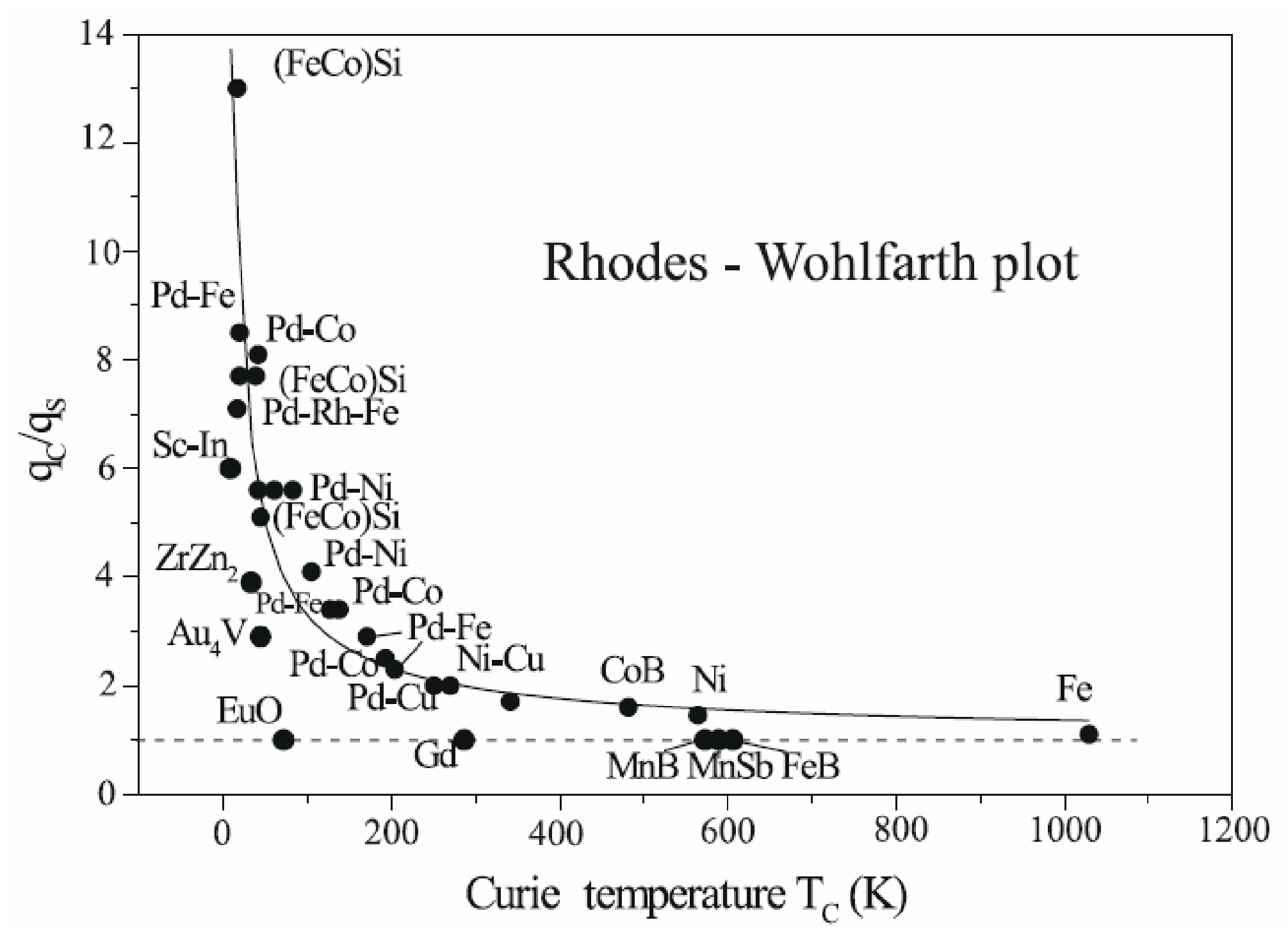

Fig. 12 The Rhodes – Wohlfarth plot. From ref. [167].

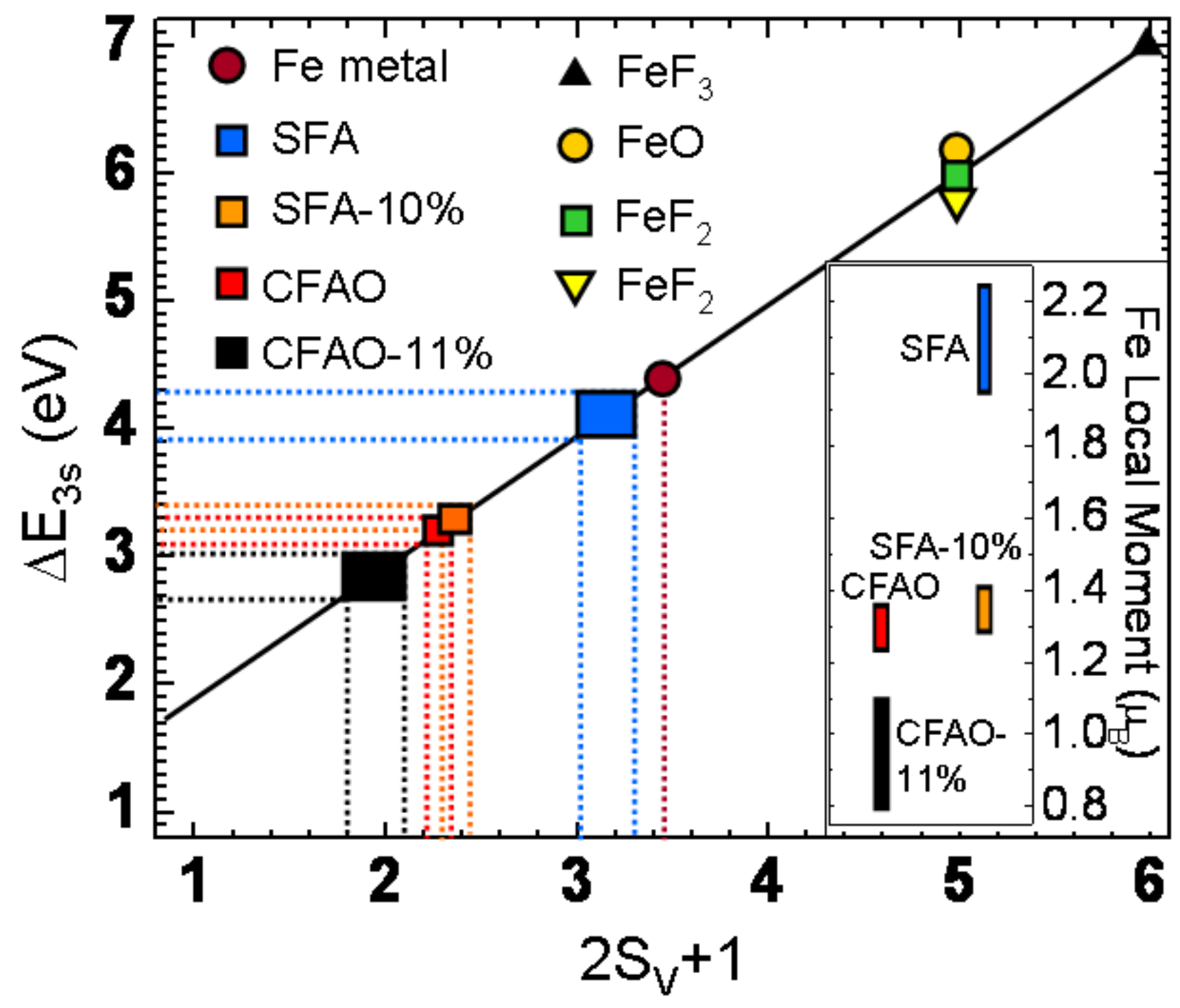

Fig. 13 Values of the spin moment on the Fe sites (inset) extracted from the multiplet energy separation $\Delta E_{3s}$. Values of the multiplet energy separation $\Delta E_{3s}$ are shown for CeFeAsO (CFAO), $CeFeAsO_{0.89}F_{0.11}$ (CFAO-11%), $SrFe_2As_2$ (SFA) and $Sr(Co_{0.12}Fe_{0.88})_2As_2$ (SFA-10%) at different temperatures and phases. The continuous line is the extrapolation of the linear fit of the $\Delta E_{3s}$ values plotted against *$(2S_V+1)$* for the Fe ionic compounds $FeF_3$, $FeF_2$, FeO, for which $S_v$ is known to be 5/2 ($FeF_3$) and 2 ($FeF_2$, FeO). The size of the symbol is much bigger than the experimental uncertainties: It denotes the range of values for the splitting $\Delta E_{3s}$, the correspondent values for $S_V$, and the Fe SM as shown in the inset. From ref. [160].

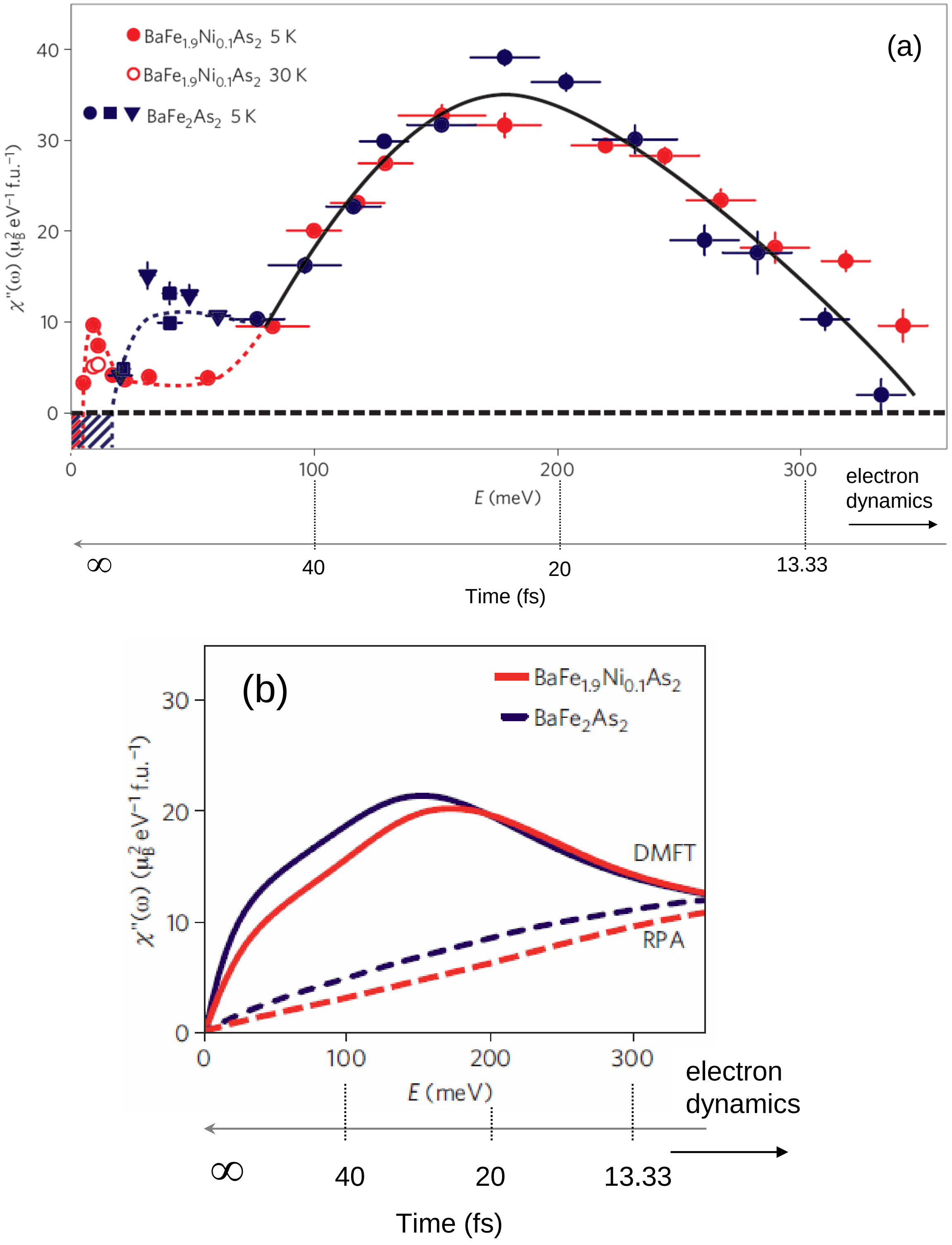

Fig. 14 (a) Dynamical spin susceptibility for $BaFe_2As_2$ and $BaFe_{1.9}Ni_{0.1}As_2$. The continuous and dotted lines are guides to the eye. (b) Calculations in absolute units of the dynamic spin susceptibility shown in (a) with DMFT (continuous line) and RPA (dashed lines). Adapted from ref. [159]. An energy integration of the function $\chi''(E)$ up to a certain value on the energy axis provides an estimate of the rapidity of the response of the system, as indicated by the values reported on the time axis.